\documentclass[twocolumn]{aastex62}

\usepackage[modulo]{lineno}
\usepackage{verbatim, graphicx,  ifthen, pbox}
\usepackage{eso-pic}
\usepackage{CJKutf8}
\usepackage{xcolor}
\usepackage{amsmath}
\usepackage{graphicx}
\usepackage{verbatim}
\usepackage{pdfpages}
\usepackage{booktabs}
\usepackage{longtable}

\makeatletter
\AtBeginDocument{\let\LS@rot\@undefined}
\makeatother

\begin{document}

\title{DYNAMICAL CLASSIFICATION OF TRANS-NEPTUNIAN OBJECTS DETECTED BY THE DARK ENERGY SURVEY}
\author[0000-0001-7721-6457]{T.~Khain}
\affiliation{Department of Physics, University of Michigan, Ann Arbor, MI 48109, USA}
\author[0000-0002-7733-4522]{J.C.~Becker}
\affiliation{Department of Physics, University of Michigan, Ann Arbor, MI 48109, USA}
\author[0000-0001-7737-6784]{Hsing~Wen~Lin (\begin{CJK*}{UTF8}{bkai}
林省文\end{CJK*})}
\affiliation{Department of Physics, University of Michigan, Ann Arbor, MI 48109, USA}
\author[0000-0001-6942-2736]{D.~W.~Gerdes}
\affiliation{Department of Astronomy, University of Michigan, Ann Arbor, MI 48109, USA}
\affiliation{Department of Physics, University of Michigan, Ann Arbor, MI 48109, USA}
\author[0000-0002-8167-1767]{F.~C.~Adams}
\affiliation{Department of Physics, University of Michigan, Ann Arbor, MI 48109, USA}
\affiliation{Department of Astronomy, University of Michigan, Ann Arbor, MI 48109, USA}
\author[0000-0003-0743-9422]{P.~Bernardinelli}
\affiliation{Department of Physics and Astronomy, University of Pennsylvania, Philadelphia, PA 19104, USA}
\author{G.~M.~Bernstein}
\affiliation{Department of Physics and Astronomy, University of Pennsylvania, Philadelphia, PA 19104, USA}
\author[0000-0002-8906-2835]{K.~Franson}
\affiliation{Department of Physics, University of Michigan, Ann Arbor, MI 48109, USA}
\author[0000-0002-2486-1118]{L.~Markwardt}
\affiliation{Department of Physics, University of Michigan, Ann Arbor, MI 48109, USA}
\author[0000-0002-6126-8487]{S.~Hamilton}
\affiliation{Department of Physics, University of Michigan, Ann Arbor, MI 48109, USA}
\author{K.~Napier}
\affiliation{Department of Physics, University of Michigan, Ann Arbor, MI 48109, USA}
\author[0000-0003-2764-7093]{M.~Sako}
\affiliation{Department of Physics and Astronomy, University of Pennsylvania, Philadelphia, PA 19104, USA}
\author{T.~M.~C.~Abbott}
\affiliation{Cerro Tololo Inter-American Observatory, National Optical Astronomy Observatory, Casilla 603, La Serena, Chile}
\author{S.~Avila}
\affiliation{Instituto de Fisica Teorica UAM/CSIC, Universidad Autonoma de Madrid, 28049 Madrid, Spain}
\author{E.~Bertin}
\affiliation{CNRS, UMR 7095, Institut d'Astrophysique de Paris, F-75014, Paris, France}
\affiliation{Sorbonne Universit\'es, UPMC Univ Paris 06, UMR 7095, Institut d'Astrophysique de Paris, F-75014, Paris, France}
\author{D.~Brooks}
\affiliation{Department of Physics \& Astronomy, University College London, Gower Street, London, WC1E 6BT, UK}
\author{E.~Buckley-Geer}
\affiliation{Fermi National Accelerator Laboratory, P. O. Box 500, Batavia, IL 60510, USA}
\author{D.~L.~Burke}
\affiliation{Kavli Institute for Particle Astrophysics \& Cosmology, P. O. Box 2450, Stanford University, Stanford, CA 94305, USA}
\affiliation{SLAC National Accelerator Laboratory, Menlo Park, CA 94025, USA}
\author{A.~Carnero~Rosell}
\affiliation{Centro de Investigaciones Energ\'eticas, Medioambientales y Tecnol\'ogicas (CIEMAT), Madrid, Spain}
\affiliation{Laborat\'orio Interinstitucional de e-Astronomia - LIneA, Rua Gal. Jos\'e Cristino 77, Rio de Janeiro, RJ - 20921-400, Brazil}
\author{M.~Carrasco~Kind}
\affiliation{Department of Astronomy, University of Illinois at Urbana-Champaign, 1002 W. Green Street, Urbana, IL 61801, USA}
\affiliation{National Center for Supercomputing Applications, 1205 West Clark St., Urbana, IL 61801, USA}
\author{J.~Carretero}
\affiliation{Institut de F\'{\i}sica d'Altes Energies (IFAE), The Barcelona Institute of Science and Technology, Campus UAB, 08193 Bellaterra (Barcelona) Spain}
\author{L.~N.~da Costa}
\affiliation{Laborat\'orio Interinstitucional de e-Astronomia - LIneA, Rua Gal. Jos\'e Cristino 77, Rio de Janeiro, RJ - 20921-400, Brazil}
\affiliation{Observat\'orio Nacional, Rua Gal. Jos\'e Cristino 77, Rio de Janeiro, RJ - 20921-400, Brazil}
\author{J.~De~Vicente}
\affiliation{Centro de Investigaciones Energ\'eticas, Medioambientales y Tecnol\'ogicas (CIEMAT), Madrid, Spain}
\author{S.~Desai}
\affiliation{Department of Physics, IIT Hyderabad, Kandi, Telangana 502285, India}
\author{H.~T.~Diehl}
\affiliation{Fermi National Accelerator Laboratory, P. O. Box 500, Batavia, IL 60510, USA}
\author{P.~Doel}
\affiliation{Department of Physics \& Astronomy, University College London, Gower Street, London, WC1E 6BT, UK}
\author{B.~Flaugher}
\affiliation{Fermi National Accelerator Laboratory, P. O. Box 500, Batavia, IL 60510, USA}
\author{J.~Frieman}
\affiliation{Fermi National Accelerator Laboratory, P. O. Box 500, Batavia, IL 60510, USA}
\affiliation{Kavli Institute for Cosmological Physics, University of Chicago, Chicago, IL 60637, USA}
\author{J.~Garc\'ia-Bellido}
\affiliation{Instituto de Fisica Teorica UAM/CSIC, Universidad Autonoma de Madrid, 28049 Madrid, Spain}
\author{E.~Gaztanaga}
\affiliation{Institut d'Estudis Espacials de Catalunya (IEEC), 08034 Barcelona, Spain}
\affiliation{Institute of Space Sciences (ICE, CSIC),  Campus UAB, Carrer de Can Magrans, s/n,  08193 Barcelona, Spain}
\author{D.~Gruen}
\affiliation{Department of Physics, Stanford University, 382 Via Pueblo Mall, Stanford, CA 94305, USA}
\affiliation{Kavli Institute for Particle Astrophysics \& Cosmology, P. O. Box 2450, Stanford University, Stanford, CA 94305, USA}
\affiliation{SLAC National Accelerator Laboratory, Menlo Park, CA 94025, USA}
\author{R.~A.~Gruendl}
\affiliation{Department of Astronomy, University of Illinois at Urbana-Champaign, 1002 W. Green Street, Urbana, IL 61801, USA}
\affiliation{National Center for Supercomputing Applications, 1205 West Clark St., Urbana, IL 61801, USA}
\author{J.~Gschwend}
\affiliation{Laborat\'orio Interinstitucional de e-Astronomia - LIneA, Rua Gal. Jos\'e Cristino 77, Rio de Janeiro, RJ - 20921-400, Brazil}
\affiliation{Observat\'orio Nacional, Rua Gal. Jos\'e Cristino 77, Rio de Janeiro, RJ - 20921-400, Brazil}
\author{G.~Gutierrez}
\affiliation{Fermi National Accelerator Laboratory, P. O. Box 500, Batavia, IL 60510, USA}
\author{D.~L.~Hollowood}
\affiliation{Santa Cruz Institute for Particle Physics, Santa Cruz, CA 95064, USA}
\author{K.~Honscheid}
\affiliation{Center for Cosmology and Astro-Particle Physics, The Ohio State University, Columbus, OH 43210, USA}
\affiliation{Department of Physics, The Ohio State University, Columbus, OH 43210, USA}
\author{D.~J.~James}
\affiliation{Center for Astrophysics $\vert$ Harvard \& Smithsonian, 60 Garden Street, Cambridge, MA 02138, USA}
\author{N.~Kuropatkin}
\affiliation{Fermi National Accelerator Laboratory, P. O. Box 500, Batavia, IL 60510, USA}
\author{M.~A.~G.~Maia}
\affiliation{Laborat\'orio Interinstitucional de e-Astronomia - LIneA, Rua Gal. Jos\'e Cristino 77, Rio de Janeiro, RJ - 20921-400, Brazil}
\affiliation{Observat\'orio Nacional, Rua Gal. Jos\'e Cristino 77, Rio de Janeiro, RJ - 20921-400, Brazil}
\author{J.~L.~Marshall}
\affiliation{George P. and Cynthia Woods Mitchell Institute for Fundamental Physics and Astronomy, and Department of Physics and Astronomy, Texas A\&M University, College Station, TX 77843,  USA}
\author{F.~Menanteau}
\affiliation{Department of Astronomy, University of Illinois at Urbana-Champaign, 1002 W. Green Street, Urbana, IL 61801, USA}
\affiliation{National Center for Supercomputing Applications, 1205 West Clark St., Urbana, IL 61801, USA}
\author{C.~J.~Miller}
\affiliation{Department of Astronomy, University of Michigan, Ann Arbor, MI 48109, USA}
\affiliation{Department of Physics, University of Michigan, Ann Arbor, MI 48109, USA}
\author{R.~Miquel}
\affiliation{Instituci\'o Catalana de Recerca i Estudis Avan\c{c}ats, E-08010 Barcelona, Spain}
\affiliation{Institut de F\'{\i}sica d'Altes Energies (IFAE), The Barcelona Institute of Science and Technology, Campus UAB, 08193 Bellaterra (Barcelona) Spain}
\author{A.~A.~Plazas}
\affiliation{Department of Astrophysical Sciences, Princeton University, Peyton Hall, Princeton, NJ 08544, USA}
\author{E.~Sanchez}
\affiliation{Centro de Investigaciones Energ\'eticas, Medioambientales y Tecnol\'ogicas (CIEMAT), Madrid, Spain}
\author{V.~Scarpine}
\affiliation{Fermi National Accelerator Laboratory, P. O. Box 500, Batavia, IL 60510, USA}
\author{M.~Schubnell}
\affiliation{Department of Physics, University of Michigan, Ann Arbor, MI 48109, USA}
\author{I.~Sevilla-Noarbe}
\affiliation{Centro de Investigaciones Energ\'eticas, Medioambientales y Tecnol\'ogicas (CIEMAT), Madrid, Spain}
\author{M.~Smith}
\affiliation{School of Physics and Astronomy, University of Southampton,  Southampton, SO17 1BJ, UK}
\author{F.~Sobreira}
\affiliation{Instituto de F\'isica Gleb Wataghin, Universidade Estadual de Campinas, 13083-859, Campinas, SP, Brazil}
\affiliation{Laborat\'orio Interinstitucional de e-Astronomia - LIneA, Rua Gal. Jos\'e Cristino 77, Rio de Janeiro, RJ - 20921-400, Brazil}
\author{E.~Suchyta}
\affiliation{Computer Science and Mathematics Division, Oak Ridge National Laboratory, Oak Ridge, TN 37831}
\author{M.~E.~C.~Swanson}
\affiliation{National Center for Supercomputing Applications, 1205 West Clark St., Urbana, IL 61801, USA}
\author{G.~Tarle}
\affiliation{Department of Physics, University of Michigan, Ann Arbor, MI 48109, USA}
\author{A.~R.~Walker}
\affiliation{Cerro Tololo Inter-American Observatory, National Optical Astronomy Observatory, Casilla 603, La Serena, Chile}
\author{W.~Wester}
\affiliation{Fermi National Accelerator Laboratory, P. O. Box 500, Batavia, IL 60510, USA}

\collaboration{The Dark Energy Survey Collaboration}
\correspondingauthor{Tali Khain}
\email{talikh@umich.edu}

\bigskip
\bigskip

\begin{abstract}

The outer Solar System contains a large number of small bodies (known as trans-Neptunian objects or TNOs) that exhibit diverse types of dynamical behavior. The classification of bodies in this distant region into dynamical classes -- sub-populations that experience similar orbital evolution -- aids in our understanding of the structure and formation of the Solar System. In this work, we propose an updated dynamical classification scheme for the outer Solar System. This approach includes the construction of a new (automated) method for identifying mean-motion resonances. We apply this algorithm to the current dataset of TNOs observed by the Dark Energy Survey (DES) and present a working classification for all of the DES TNOs detected to date. Our classification scheme yields 1 inner centaur, 19 outer centaurs, 21 scattering disk objects, 47 detached TNOs, 48 securely resonant objects, 7 resonant candidates, and 97 classical belt objects. Among the scattering and detached objects, we detect 8 TNOs with semi-major axes greater than 150 AU. \\
\end{abstract}

\section{Introduction} \label{sec:intro}

Our Solar System harbors a large collection of small icy bodies that orbit the Sun beyond Neptune. In the past two decades, the number of these trans-Neptunian objects (TNOs) that has been discovered has grown to thousands. As these objects are believed to be primordial tracers of the early Solar System, the characterization of the trans-Neptunian population is vital for understanding and testing theoretical models of Solar System formation. For example, in one class of theories collectively known as the Nice Model \citep{2005Natur.435..459T,2011ApJ...742L..22N,2012ApJ...744L...3B}, the starting orbits of the giant planets are different from those of the present epoch. Such models predict sizes and distributions of the different sub-populations of TNOs in the Kuiper belt due to the orbital migration of the larger planets to their current locations. 

Over the past decades, a number of surveys intended to study the outer Solar System have significantly increased the population of known TNOs \citep[e.g.,][]{2001AJ....122..457T, 2014AJ....148...55A, 2010ApJ...720.1691S, 2011AJ....142..131P, 2018ApJS..236...18B}, allowing these theories to be tested. Today, the growing number of observed objects combined with the development of survey simulators \citep{2018FrASS...5...14L, Stephanie} allows for detailed comparisons of the observed and predicted populations \citep{2016AJ....152...23V, 2018AJ....155..260V} as expected within single modern surveys.

The trans-Neptunian objects themselves can be characterized in a variety of ways, including their size, color, and composition. These physical properties of the objects, however, are often difficult to observe. Fortunately, the orbits of the objects can provide insight into the structure and dynamical history of this distant region. By categorizing the TNOs based on their dynamical behaviors, we can extract information about the various sub-populations of the outer Solar System. The primary works that laid out this type of dynamical classification scheme are those of \citet{Elliot2005} and \citet{2008ssbn.book...43G}; the major dynamical classes of the Kuiper belt include the Neptune-resonant objects, centaurs, scattering disk objects, detached TNOs, and more (see below).

One of the surveys that has led to the discovery of these Kuiper belt objects is the Dark Energy Survey (DES) \citep{des_survey}, a nominal five year baseline optical survey intended primarily for cosmological purposes.
DES used the Dark Energy Camera \citep[DECam,][]{2015AJ....150..150F} on the 4-meter Blanco telescope at the Cerro Tololo Inter-American Observatory in Chile. 
Its survey area subtended a total of 5000 square degrees of sky, which was tiled with two survey modes: the Wide Survey, which imaged the full survey area roughly twice per year to a limiting magnitude of $r\sim23.8$ mag for single epoch exposures in each of the \emph{grizY} bands; and the Supernova Survey \citep{2012ApJ...753..152B}, which consisted of 30 square degrees spread over ten regions, each of which were imaged roughly weekly in the \emph{griz} bands. 

In a partial search of its first four years of data, DES has detected over two hundred TNOs (and counting). The discoveries so far include Neptune trojans \citep{2016AJ....151...39G, 2019Icar..321..426L}, a dwarf planet candidate \citep{2017ApJ...839L..15G}, two members of a potentially associated triplet family \citep{2018AJ....156..273K}, and a high-inclination extreme TNO \citep{2018AJ....156...81B}, with further publications detailing the results of additional analysis to come. Now that the current DES dataset has grown to this substantial size, it is of great interest to study the dynamical properties of this TNO population.

In this work, we present the dynamical classification of 240 trans-Neptunian objects detected by the Dark Energy Survey. Although the present application is to this particular set of TNOs, the classification scheme developed herein can be used more broadly. In Section \ref{sec:method}, we lay out the different categories of TNOs and our classification algorithm, which differs somewhat from that of \citet{2008ssbn.book...43G}. In addition, we outline our newly developed resonance-finding method that allows for an automated resonance search without visual inspection. In Section \ref{sec:results}, we apply this algorithm to the object sample and present the classification of the known DES TNOs. We discuss our results and their implications for future work in Section \ref{sec:discussion}.

\begin{figure}[t!]
\epsscale{1}
  \begin{center}
      \leavevmode
\includegraphics[width=85mm]{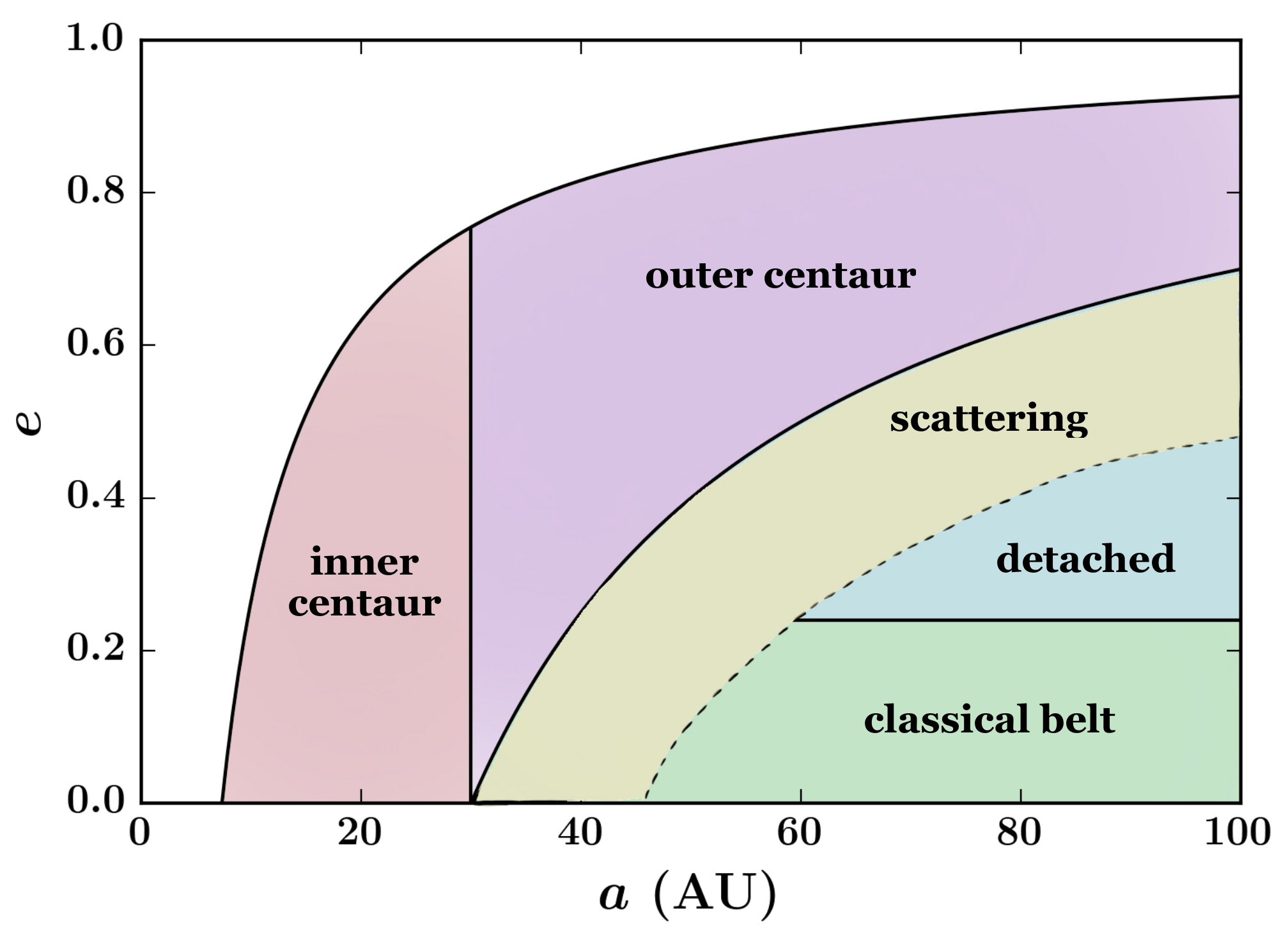}
\caption{The dynamical classes of the outer Solar System. The black solid curves correspond to constant perihelion distances, with $q = 7.35$ AU and $q = 30$ AU (top to bottom). The inner centaurs (red region) have orbital periods less than Neptune's. The outer centaurs (purple) have orbits with perihelion distances below Neptune's orbit, but with semi-major axes outside the giant planet region. The scattering population (SDOs, scattering disk objects) mostly lies along the $q = 30$ AU curve and is shown in yellow. The classical belt (green region) and the detached objects (blue region) are removed from the Neptune scattering region, with the higher eccentricity detached TNOs above the classical belt. A companion plot with the DES TNOs on this phase plane is found in Figure \ref{fig:ae_plane}.}
\label{fig:ae_plane_colors}
\end{center}
\end{figure}

\section{Classification Method} \label{sec:method}

In this work, we apply the classification scheme of \citet{2008ssbn.book...43G} with a few changes that reflect the development of the field in the last decade. The categories of objects and the definitions we adapt are described below and are visually represented in Figure \ref{fig:ae_plane_colors}. As with any classification scheme, a few of the category boundaries are rather arbitrary, as some of these dynamical properties lie on a spectrum.
Deviations from \citet{2008ssbn.book...43G} are denoted with an asterisk$^*$.

\textbf{Jupiter-coupled object.}
Jupiter-coupled objects are defined through the Tisserand paramater $T_J$ with respect to Jupiter,

\begin{equation}
T_J = \frac{a_J}{a} + 2 \sqrt{\frac{a}{a_J}(1 - e^2)} \cos{i},
\end{equation}
where $a_J$ is the semi-major axis of Jupiter, and $a, e, i$ are the semi-major axis, eccentricity, and inclination of the object, respectively. Objects with $T_J < 3.05$ and perihelion distances below $q < 7.35$ AU are considered to be Jupiter-coupled objects. 

Since the current DES sample does not contain any objects which exhibit cometary dynamics, we drop this category in future discussion of the classification results.

\textbf{Centaur$^*$.} Centaurs are objects that experience strong interactions with the giant planets. In this work, we propose to separate this class into two: inner centaurs and outer centaurs. Inner centaurs (the traditional centaurs described in \citealt{2008ssbn.book...43G}) are objects with semi-major axes smaller than Neptune's ($a < a_N \approx 30$ AU). We define outer centaurs to be objects with perihelion distances shorter than Neptune's semi-major axis ($q < a_N$), but semi-major axes larger than Neptune's semi-major axis ($a > a_N$). 

Although both types of centaurs spend time within the giant planet region, the frequency of interactions with the planets differs for each class. The inner centaurs may experience strong interactions with the giant planets at most points on their orbit, while the outer centaurs are affected once an orbit, during perihelion crossing; moreover, the orbital period of an outer centaur is longer than of an inner centaur, resulting in fewer interactions per unit time. This distinction highlights the difference in the instability timescale: the outer centaurs are longer-lived objects than the short lifetime inner centaurs \citep{2003AJ....126.3122T, 2004MNRAS.354..798H}. By this classification, a traditional centaur such as Chiron \citep{1979IAUS...81..245K} falls into the inner centuar category, while longer-period objects with high eccentricity such as Drac \citep{2009ApJ...697L..91G} or Niku \citep{2016ApJ...827L..24C} are deemed outer centaurs.

\begin{figure*}[h!]
\epsscale{1}
  \begin{center}
      \leavevmode
\includegraphics[width=170mm]{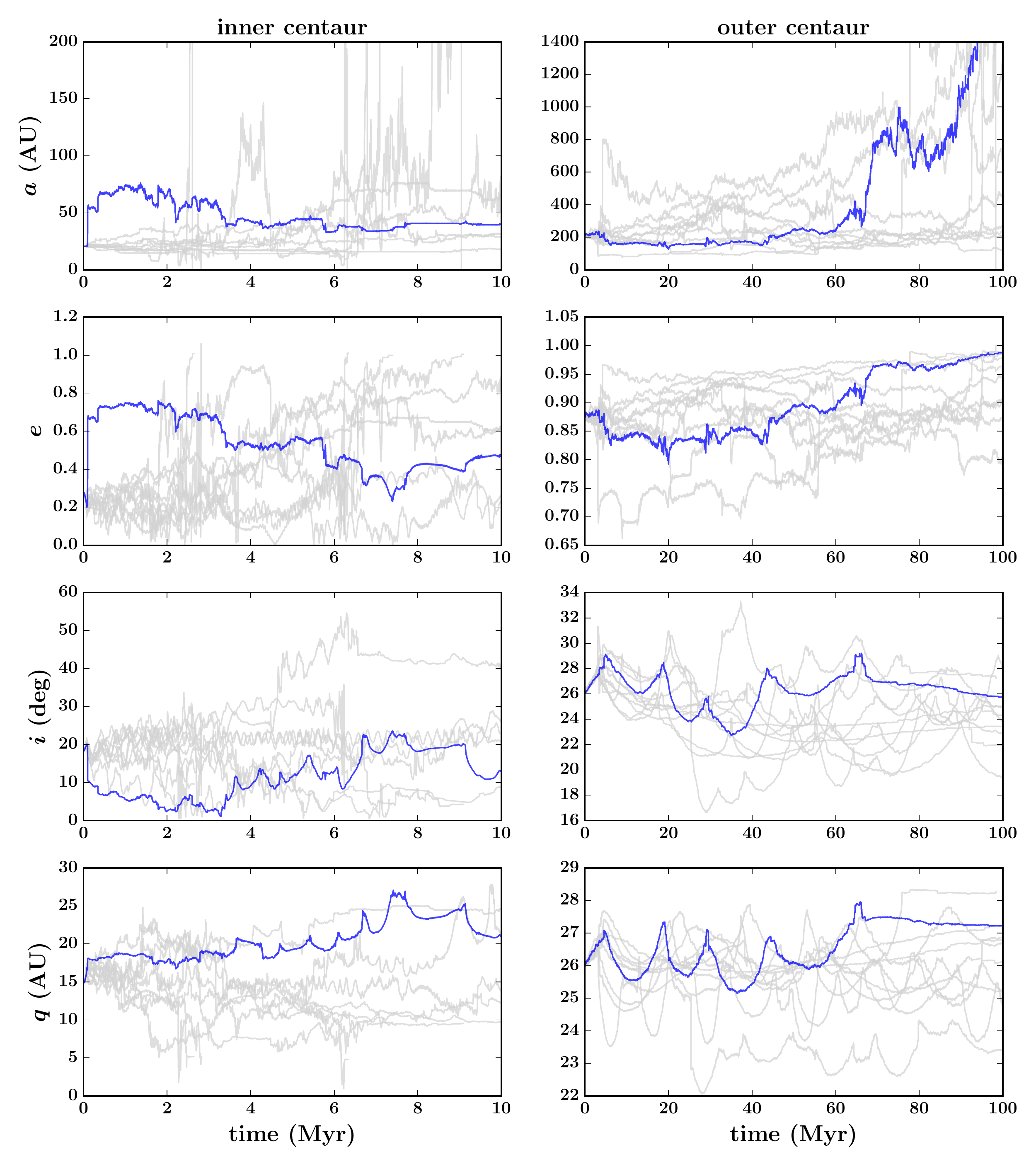}
\caption{Example dynamics of an inner centaur (left column, object 2003 QC$_{112}$) and an outer centaur (right column, object \texttt{s11\_good\_19}) detected in the DES data. The panels show the time evolution of semi-major axis, eccentricity, inclination, and perihelion distance. The trajectories of the ten clones are shown in gray and the best fit trajectory is in blue. Note the short perihelion distance of the two centaurs.}
\label{fig:tnoclass_1}
\end{center}
\end{figure*}

An example of the dynamics of inner and outer centaurs from the DES set is shown in Figure \ref{fig:tnoclass_1}.

\textbf{Oort cloud object.} Objects in the Oort cloud are defined to have semi-major axes $a > 2000$ AU. Due to their large orbits, these bodies are most likely affected by galactic tides and passing stars. The present DES sample does not contain any objects in this class. 

\textbf{Resonant object.} The outer Solar System consists of a large number of TNOs in mean motion resonances with Neptune. In order to be in a Neptune mean motion resonance, a TNO must be near an integer period ratio with Neptune's period, and must have a librating resonance argument of the form
\begin{equation}
\phi = p\lambda_N + q\lambda + r\varpi_N + s\varpi,
\end{equation}
\label{eq:resarg}
where $p, q, r$, and $s$ are integers that satisfy the d'Alembert relation, $p + q + r + s = 0$. Here, $\lambda = \Omega + 
\omega + M$ is the mean longitude, $\varpi = \Omega + 
\omega$ is the longitude of perihelion, the subscript $N$ refers to Neptune's orbital elements, and the non-subscripted variables refer to the TNO. Such a resonance is then referred to as a $p$:$q$ resonance, the ratio of Neptune's orbital period to that of the TNO. In this work, we only consider the eccentricity-type resonances given by Equation \ref{eq:resarg}, as was done in \citet{2008ssbn.book...43G}. In theory, TNOs could also experience inclination-type resonances, which include independent $\Omega$ and $\Omega_N$ terms. Since these are a higher order effect, we leave the study of inclination-type resonances for future work.

\begin{figure*}[p]
\epsscale{1}
  \begin{center}
      \leavevmode
\includegraphics[width=170mm]{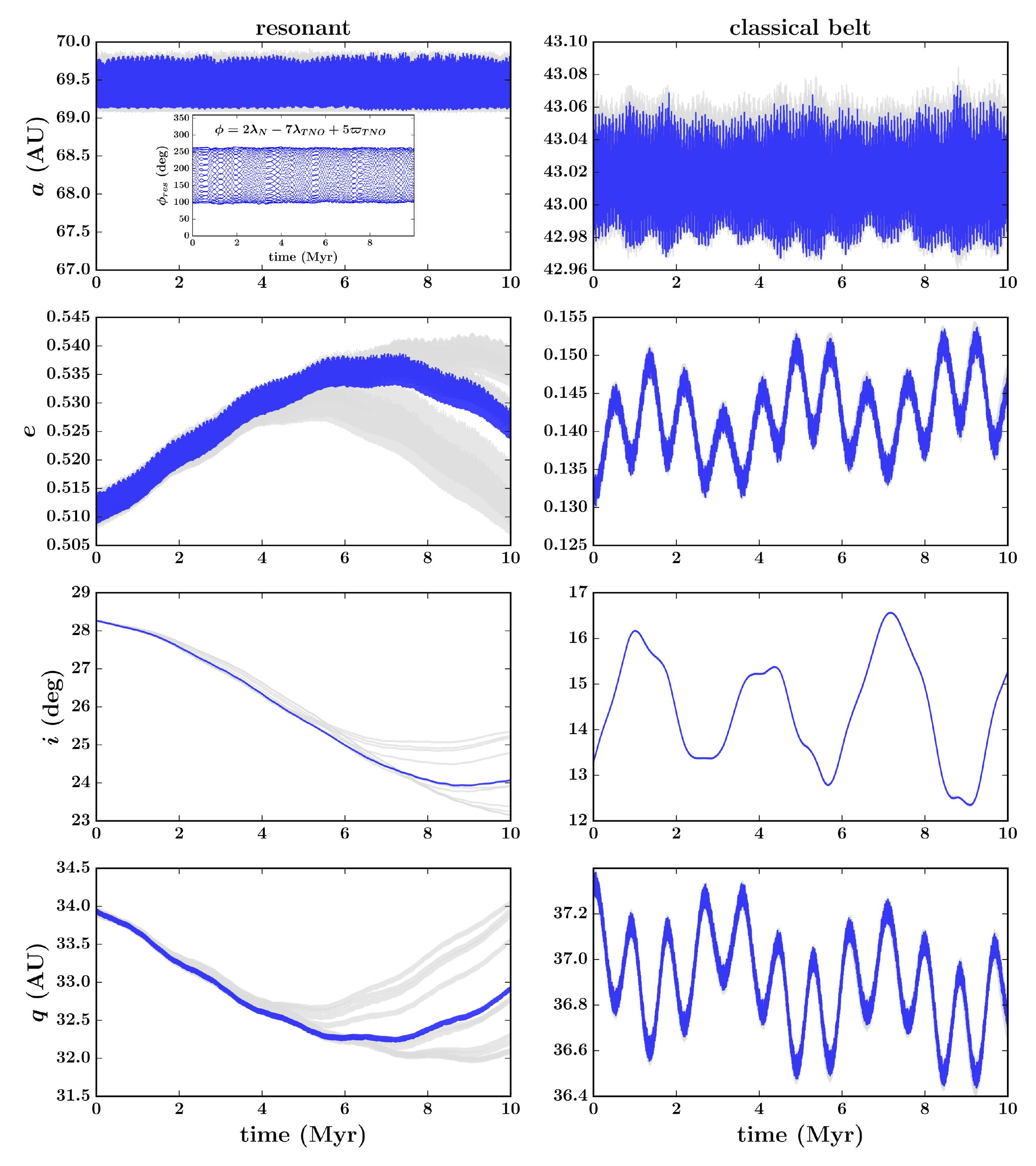}
\caption{Example dynamics of a resonant object (left column, object \texttt{s12\_good\_5}) and a classical belt object (right column, object 2013 RP$_{98}$) detected in the DES data. The panels show the time evolution of semi-major axis, eccentricity, inclination, and perihelion distance. The trajectories of the ten clones are shown in gray and the best fit trajectory is in blue. The inset in the top left panel displays the time evolution of the resonant argument corresponding to the $2$:$7$ resonance of the TNO; note that the behavior of this angle is bounded (librating), indicating that this TNO is in fact in resonance for the full integration time.}
\label{fig:tnoclass_3}
\end{center}
\end{figure*}

An example of a resonant TNO from the DES data is shown in the left column of Figure \ref{fig:tnoclass_3}. Note the constant behavior of the semi-major axis in the top panel; the inset demonstrates the librating resonance argument corresponding to the $2$:$7$ commensurability. 

\begin{figure*}[h!]
\epsscale{1}
  \begin{center}
      \leavevmode
\includegraphics[width=170mm]{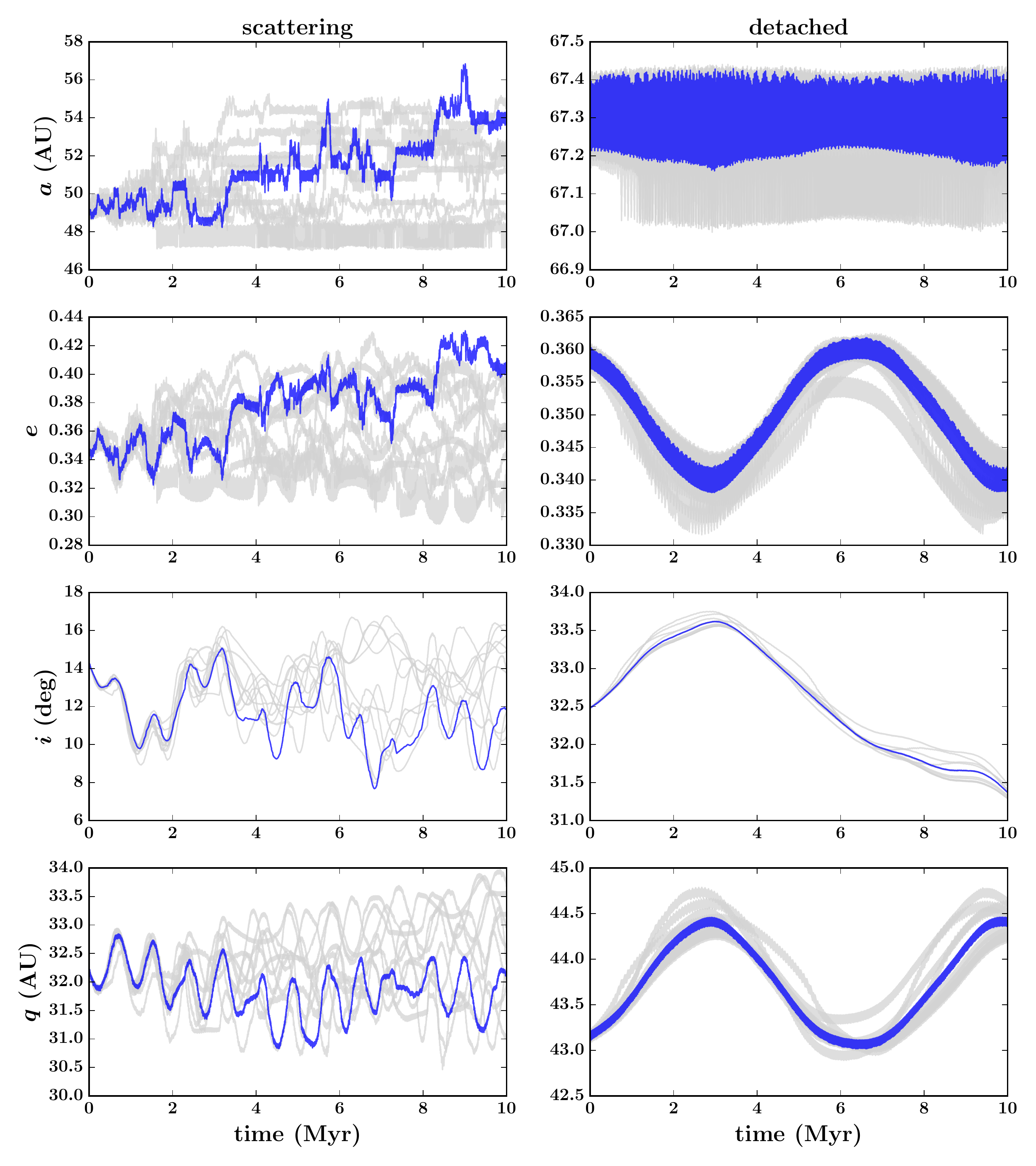}
\caption{Example dynamics of a scattering object (left column, object 2012 WG$_{37}$) and a detached object (right column, object \texttt{s14\_good\_4}) detected in the DES data. The panels show the time evolution of semi-major axis, eccentricity, inclination, and perihelion distance. The trajectories of the ten clones are shown in gray and the best fit trajectory is in blue. Note the varying semi-major axis of the scattering object (left) and the contrasting constant $a$ behavior of the detached object (right).}
\label{fig:tnoclass_2}
\end{center}
\end{figure*}

\textbf{Scattering disk object (SDO)$^*$.} SDOs are objects that are currently scattering off of Neptune, and experience rapid and significant variations in their semi-major axis evolution as a result. Unlike the outer centaurs, the orbits of the scattering objects lie fully outside the giant planet region, and thus SDOs experience rather weak interactions with Neptune. Consistent with the \citet{2008ssbn.book...43G} definition, we define a scattering object as one whose semi-major axis changes by more than a few AU with respect to its initial value, $a_0$, over the integration time (10 Myr for objects with $a < 100$ AU, and 100 Myr for objects with $a > 100$ AU). To ensure that this definition scales well as we consider longer period objects, our criterion for scattering is as follows:
\begin{equation}
\frac{\Delta a}{a} > 0.0375,
\end{equation}
where 
\begin{equation}
\frac{\Delta a}{a} = \frac{\max{(a(t) - a_0)}}{a_0}
\end{equation}
is the maximum variation in semi-major axis over the integration time. The choice in the exact value of variation allowed before an object becomes scattering is somewhat arbitrary, but must be large enough that periodic variations of orbital elements do not falsely classify an object as scattering. Here we use the value of 0.0375, as it corresponds to the accepted change of 1.5 AU for a typical classical belt object at $a = 40$ AU \citep{2008ssbn.book...43G}. Previous works have also used $\Delta a/a <0.05$ \citep{2017AJ....154...62V} and 1.5 AU \citep{2004MNRAS.355..935M}. An example of the dynamics of a scattering object from the DES set is shown in the left column of Figure \ref{fig:tnoclass_2}. Note the significant change in the semi-major axis over the short 10 Myr integration time, as well as the proximity of the perihelion distance to Neptune's orbit at 30 AU. 

\textbf{Detached object.} Detached TNOs are objects whose dynamics are decoupled from Neptune's influence. Generally, these are TNOs with large perihelion distances; following \citet{2008ssbn.book...43G}, we define non-scattering and non-resonant TNOs with eccentricities $e > 0.24$ to be detached. Most of these objects are found beyond the $1$:$2$ resonance with Neptune ($a > 47.7$ AU). An example of a detached TNO is shown in the right column of Figure \ref{fig:tnoclass_2}. Note the large perihelion distance and the resulting undisturbed semi-major axis evolution.

\textbf{Classical belt object.} The classical belt, then, is composed of non-scattering TNOs with eccentricities $e < 0.24$. An example of such an object is shown in the right column of Figure \ref{fig:tnoclass_3}. 

A visual representation of these dynamical regimes on the semi-major axis - eccentricity plane can be found in Figure \ref{fig:ae_plane_colors}. A companion plot that shows the DES TNOs in each class and a detailed discussion of these results is found in Section \ref{sec:results}.

Given the definitions above, we begin by checking each object in our sample for resonant behavior. If non-resonant, we proceed to classify its dynamics into one of the remaining classes.

Although it may be possible to determine whether an object fits into one of the above categories just by considering its present day orbit, we cannot fully classify the objects without understanding their orbital evolution. The two categories that require this knowledge are the resonant and scattering classes; without running numerical simulations that model the outer Solar System, we cannot classify such objects. 

Using the categories outlined above, we present our algorithm for TNO classification below.

\begin{enumerate}
	\item From observations, determine the best-fit orbital elements and the associated covariance matrix for each object. In this work, we use the fitting algorithm from \citet{2000AJ....120.3323B}. 
	\item Generate ten clones of each TNO by drawing from a six-dimensional Gaussian distribution, where the best-fit orbit is the mean and the covariance matrix represents the uncertainties.
	\item Run an N-body integration of the ten clones and the best-fit orbit. In order to properly compare classifications for different objects, it is best if the dynamical behavior is evaluated for approximately the same number of orbital periods. For this reason, we run 10 Myr integrations for objects with $a < 100$ AU and 100 Myr integrations for objects with $a > 100$ AU. The threshold of 100 AU is an arbitrary choice, but the integrations must be extended for longer period objects as it takes more time to evaluate the dynamics. We use the N-body code \texttt{mercury6} with a hybrid symplectic and Bulirsch-Stoer (B-S) integrator and a time step of 20 days. In each integration, we include the TNO and its clones as test particles, as well as the four giant planets as active bodies (Jupiter, Saturn, Uranus, Neptune). We integrate the orbital elements for each TNO to a common epoch before beginning the simulations; in this work, time zero corresponds to the date May 4th, 2019.
	\item Dynamically classify the objects based on the output of the simulations. The TNOs are grouped into the Jupiter-coupled object, inner centaur, outer centaur, Oort cloud, detached, and classical belt classes based on the current day best-fit orbit. The resonant and scattering classifications are based on the time-evolution of the ten clones. In particular, we consider TNOs with more than five clones that experience scattering behavior (as defined above) to be scattering objects. The resonant classification is more strict; only objects that are resonant for greater than 95$\%$ of the time, averaging over the ten clones, are considered to be resonant objects. Additional details regarding the resonance classification can be found in Section \ref{sec:res}.
	\item Check if there are objects with insecure classifications. Such TNOs generally have clones with orbits that are different enough to cause them to experience disparate dynamical evolution. For example, in our data, we found that a handful of TNOs would have a couple of scattering clones, but the rest of their clones would be detached. In this situation, we extend the integration time to 100 Myr to enable a more secure classification. If the classification remains insecure, we sort the object into a category as delineated in step 4, and leave the question of secure classification for future work, once higher precision orbits are acquired.
\end{enumerate}

As can be seen from the dynamical class definitions above, it is straightforward to automatically separate the TNOs into the Jupiter-coupled object, inner centaur, outer centaur, Oort cloud, scattering, detached, and classical belt categories. The tricky step of the process is the resonance classification. To classify an object as resonant, it must not only be near an integer period ratio with Neptune, but we must identify a librating resonance angle. Often in the literature, this analysis is done by hand. Since the DES dataset contains hundreds of objects, this becomes significantly time intensive. In addition, since each period ratio has a large number of resonance arguments associated with it (i.e. for each $p,q$ pair, there are many $r,s$ pairs that satisfy $p + q + r + s = 0$), it is difficult to conclude with certainty that an object is non-resonant. 

In the following subsection, then, we describe the resonance identification algorithm we have developed to address these challenges. The main idea behind the algorithm lies in plotting the time evolution of many potential resonance arguments, and searching for regions of libration by identifying low point density regions in the plot. By applying this strategy, we are able to successfully identify a number of resonant objects, some of which are in rather high order resonances with Neptune.

\vspace{5mm}

\subsection{Resonance Identification}\label{subsec:res_identify}

\begin{figure}[t!]
\epsscale{1}
  \begin{center}
      \leavevmode
\includegraphics[width=85mm]{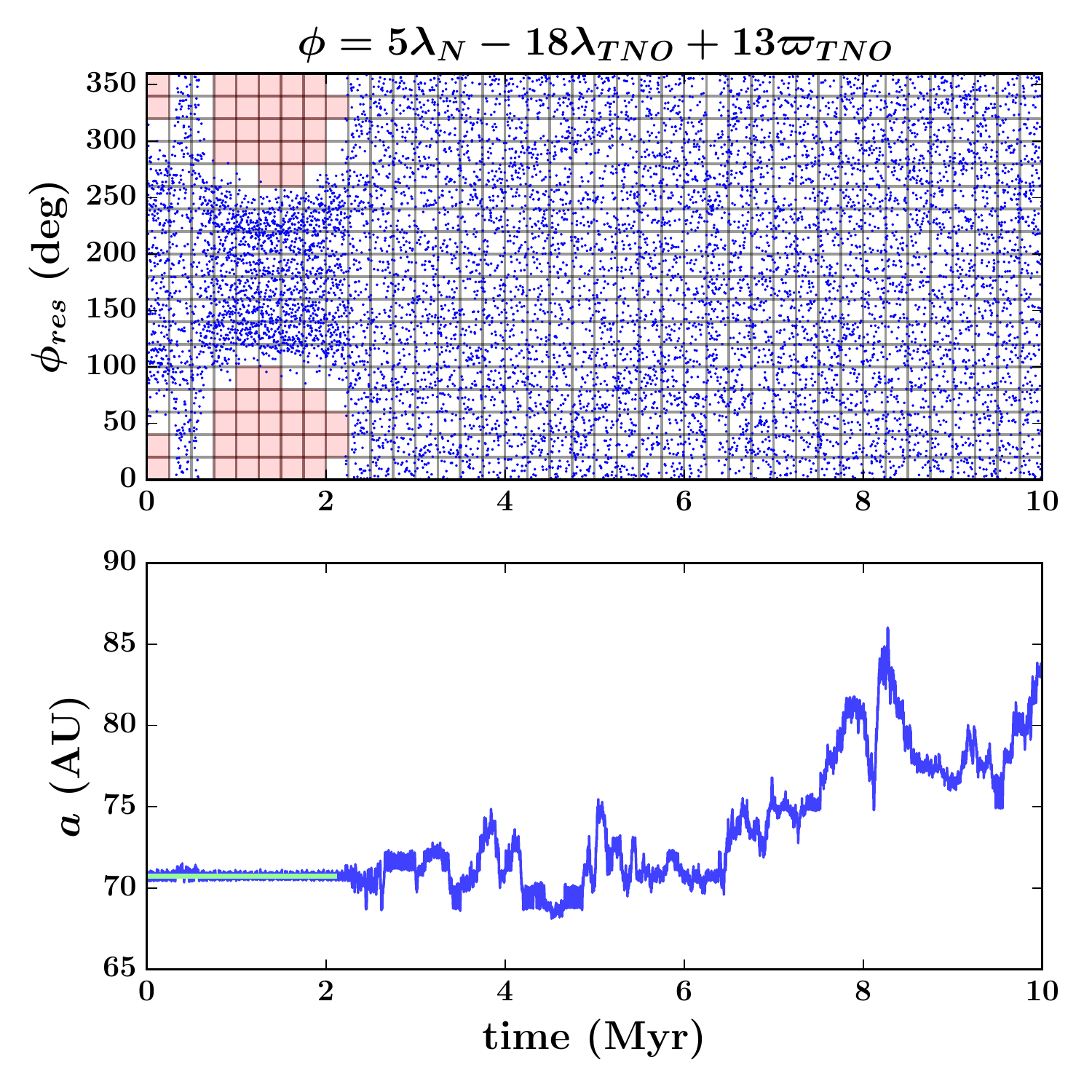}
\caption{A demonstration of the automated resonance identification algorithm. The top panel shows the time-evolution of the resonance argument $\phi$ in small blue markers. The grid guides the search for low-point-density rectangles, which are shaded in light red. The bottom panel shows the corresponding semi-major axis evolution, with regions of constant $a$ highlighted in green. Note that this figure demonstrates a likely non-resonant object; this particular clone only spends a small portion of the integration time in resonance.}
\label{fig:res_alg_bad}
\end{center}
\end{figure}

\begin{figure}[t!]
\epsscale{1}
  \begin{center}
      \leavevmode
\includegraphics[width=85mm]{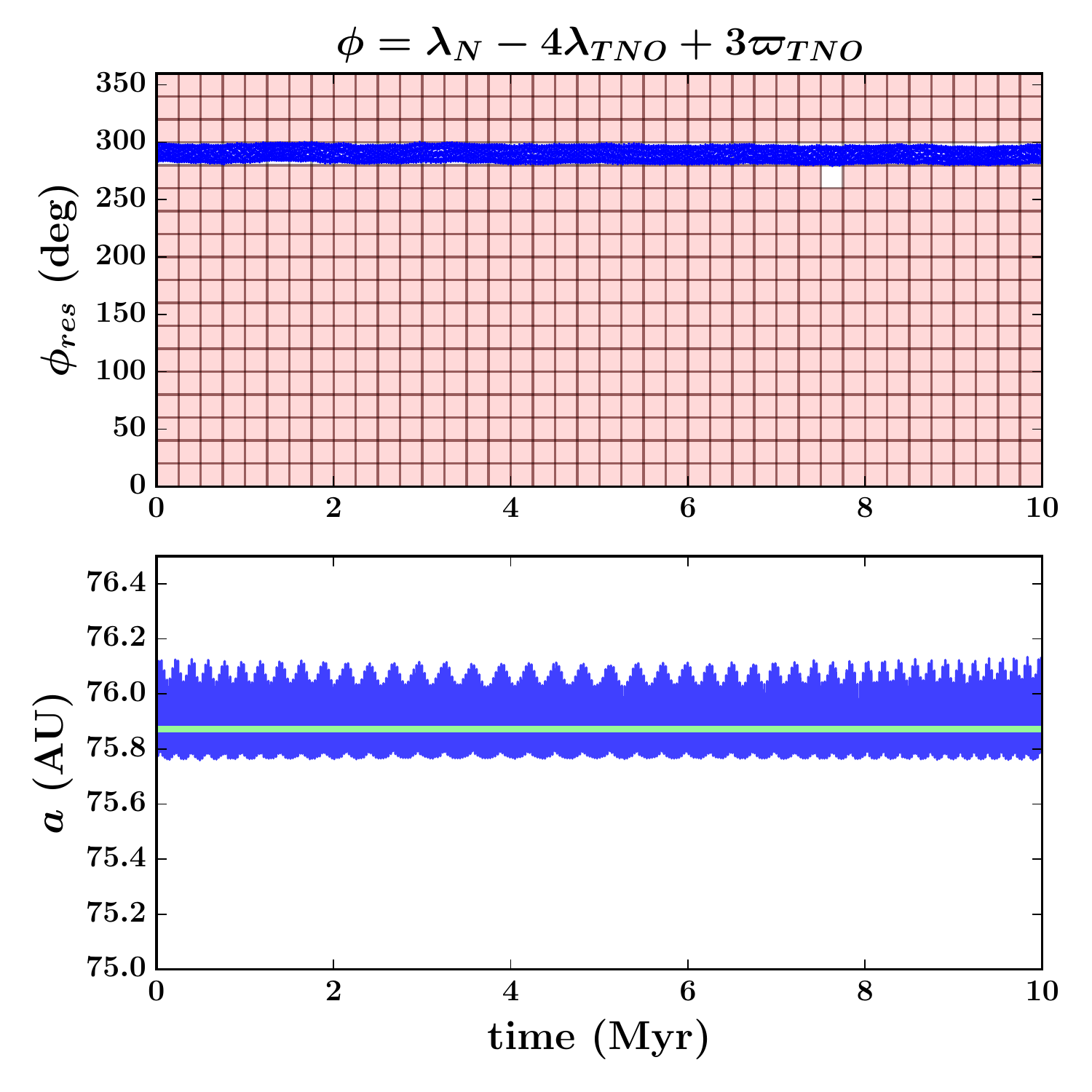}
\caption{A demonstration of the automated resonance identification algorithm. The top panel shows the time-evolution of the resonance argument $\phi$ in small blue markers. The grid guides the search for low-point-density rectangles, which are shaded in light red. The bottom panel shows the corresponding semi-major axis evolution, with regions of constant $a$ highlighted in green. In contrast to Figure \ref{fig:res_alg_bad}, this clone is in resonance for the full integration time. The large number of shaded grid squares indicate the clearly bounded resonance angle evolution.}
\label{fig:res_alg_good}
\end{center}
\end{figure}

In this subsection, we describe the resonance identification process. The input for this algorithm are the simulation results for the ten clones of the TNO; each clone is studied individually, as described below. A sample of this procedure is demonstrated in Figures \ref{fig:res_alg_bad} and \ref{fig:res_alg_good}. 

\begin{enumerate}
	\item Divide the total integration time into shorter time intervals. Since the algorithm is based on a point-density analysis, we have found that it is best if each interval contains $\sim$ 5000 data points. In our 10 Myr integrations, this corresponds to 5 Myr intervals, and 50 Myr intervals in our 100 Myr integrations. This coarse subdivision allows us to identify regions of constant semi-major axis; as described below, we break these time intervals up further in later steps of the process. 
	\item Average over the semi-major axis evolution in each interval, and compute the corresponding averaged period ratio with Neptune, $R_{av}$.
	\item If the average period ratios in neighboring intervals have similar values, connect the time intervals. In our analysis, we connect these intervals if the period ratios differ by less than 0.01. In the steps that follow, we will search for resonances in each of these connected intervals.
	\item Recall that the resonance argument is of the form

    \begin{equation}
    \phi = p\lambda_N + q\lambda + r\varpi_N + s\varpi,
    \end{equation}
    where $p, q, r$, and $s$ are integers that satisfy $p + q + r + s = 0$. For each interval, consider a range of $p$:$q$ resonances that span the period ratio range of ($R_{av}$ - resonance width, $R_{av}$ + resonance width). In our analysis, we use a resonance width value of 0.2, which corresponds to a range of about 7 AU at a semi-major axis of 39 AU. Note that this purposefully overestimates the resonant width to ensure that all possible resonances are considered; realistic calculations of the semi-major axis width for Neptune resonances can be found in \citet{2017AJ....154...20W, 2019arXiv190106040L}.
    
    \item Identify the first $p$:$q$ resonance within the period ratio range. Here, a decision needs to be made regarding the order of the resonances considered. In our analysis, we check all resonance arguments with $p, |q| \in [1, 26]$, and $r, s \in [-25, 24]$. 
    \item Fix the first pair of $r$ and $s$ coefficients.
    \item Next, overlay a fine grid on the plot of $\phi$ vs. time over one time interval. We use a grid of 18 horizontal lines, as $\phi \in (0^{\circ}, 360^{\circ})$, and 20 vertical lines for every 5000 points (see top panel of Figures \ref{fig:res_alg_bad}-\ref{fig:res_alg_good}).
    \item Run over the grid, counting the number of points in each grid square. Flag grid squares with few points (for the parameters specified above, we flag squares with one or zero points). In Figure \ref{fig:res_alg_bad}-\ref{fig:res_alg_good}, flagged squares are shaded in light red. Next, impose additional restrictions on the grid to correctly identify resonances; we require that there must be at least two flagged squares per column, or at least two adjacent flagged squares per row, and require a total number of flagged squares to exceed a set threshold. These additional conditions help discard false positives, and can be adjusted depending on the data one is working with.
    \item Repeat steps 6-8 for each pair of $r, s$ coefficients which satisfy the resonance relationship for the chosen $p$:$q$ resonance. Once all $r, s$ pairs have been cycled through, identify the best $r, s$ pair by choosing the one with the largest number of flagged grid squares.
    \item Repeat steps 5-9 for the entire set of $p$:$q$ pairs.
    \item Repeat steps 1-10 for each clone of the TNO. Compute the fraction of time spent in resonance by each clone, and average over all clones to find the resonance percentage for the TNO. 
\end{enumerate}

In this process, then, we parse the simulation data on a variety of timescales. First, we identify the regions of constant semi-major axis on long time intervals, and then check the resonance argument libration precisely on a fine subdivided grid. To achieve the best results, the exact length of these intervals should scale with the orbital period of the object one is studying.

After applying this algorithm, a decision needs to be made regarding the percentage threshold at which a TNO is considered to be truly \textit{resonant}. In our analysis, we define objects that are resonant for greater than 95$\%$ of the time to be resonant, and objects that are resonant for greater than $50\%$ of the time to be resonant candidates. The application of this procedure to the current DES TNO sample and the analysis of the results is described in the following section. 

\section{Classification of DES TNOs} \label{sec:results}

We apply the algorithm described in Section \ref{sec:method} to the currently available dataset of DES TNOs. The sample does not contain any Jupiter-coupled objects or Oort cloud objects, but all other dynamical classes are represented. We find 1 inner centaur, 19 outer centaurs, 21 scattering disk objects, 47 detached TNOs, 48 securely resonant objects, 7 resonant candidates, and 97 classical belt objects.
The classifications for specific objects and their barycentric orbital elements are reported in \ref{tab:tno_data}.

\begin{figure}[h!]
\epsscale{1}
  \begin{center}
      \leavevmode
\includegraphics[width=85mm]{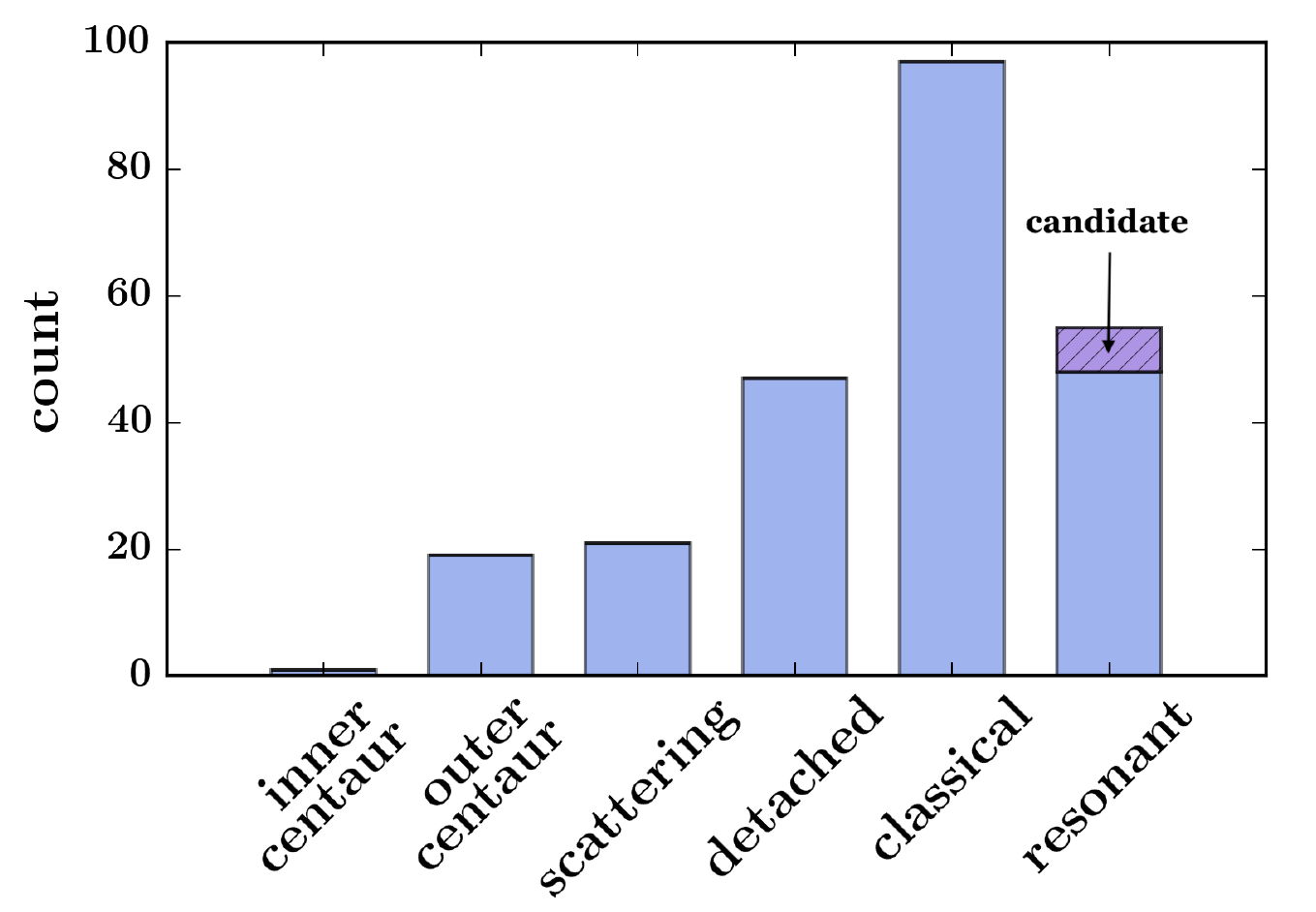}
\caption{A summary plot of the dynamical classification of the DES TNOs, showing the relative abundance of each category out of the 240 total classified objects. Most of the objects detected in the data are members of the classical belt, but there are number of both detached and securely resonant objects as well. Resonant objects that could not be securely identified are marked as candidates.}
\label{fig:tno_class}
\end{center}
\end{figure}

A visual summary of these results is shown in the bar plot in Figure \ref{fig:tno_class}. The classical belt population dominates the dataset, but there is a significant number of detached and resonant TNOs as well. The resonant bar consists of two parts; the blue represents the securely resonant objects, while the purple shows the resonant candidates. 

\begin{figure*}[t!]
\epsscale{1}
  \begin{center}
      \leavevmode
\includegraphics[width=176mm]{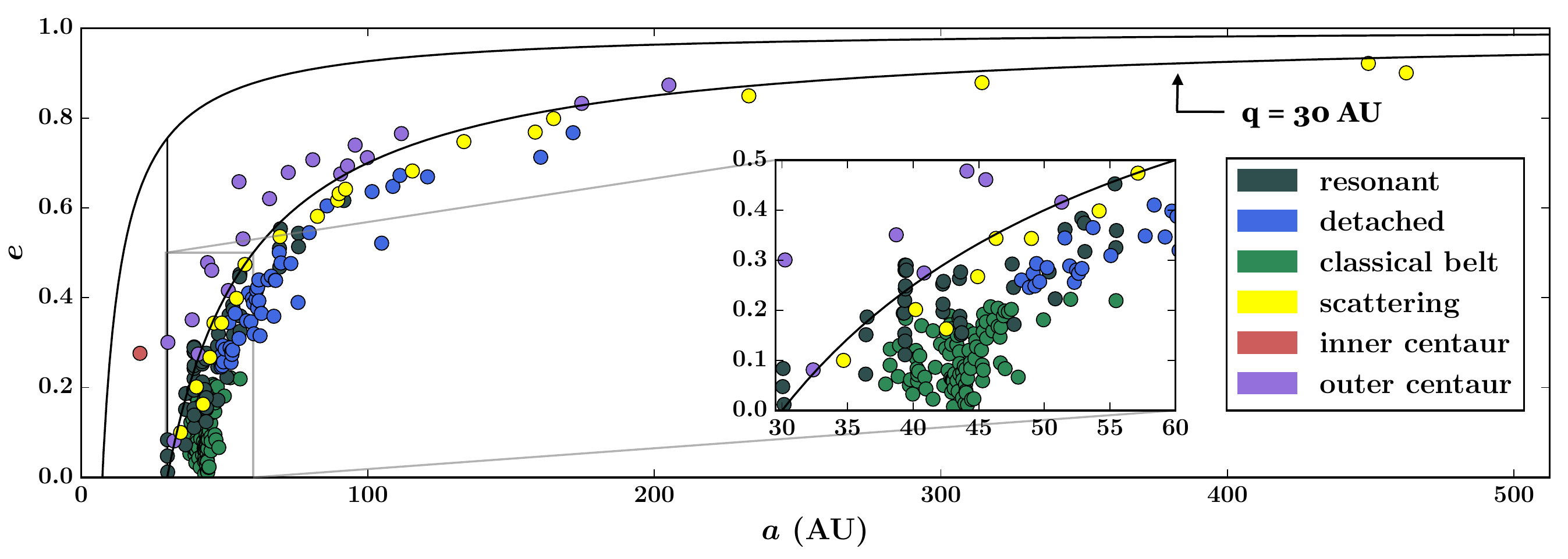}
\caption{The DES TNOs on the semi-major axis-eccentricity plane, with colored markers indicating the different dynamical classes into which objects have been classified. The black solid curves correspond to constant perihelion distances, with $q = 7.35$ AU and $q = 30$ AU (top to bottom); detections are biased towards objects whose current distances are closer, leading the envelope of the largest density of discovered objects to have a rough outer limit at around $q = 35 - 36$ AU. A companion plot that denotes the approximate region of each dynamical class is found in Figure \ref{fig:ae_plane_colors}.  The inset zooms in on the $a \in [30, 60]$ AU, $e \in [0, 0.5]$ region of the outer Solar System. The orbital elements of the objects are plotted at the epoch reported in Table \ref{tab:tno_data}.}
\label{fig:ae_plane}
\end{center}
\end{figure*}

This data is further visualized on the semi-major axis-eccentricity plane in Figure \ref{fig:ae_plane}. The black solid curves correspond to constant perihelion distances, with $q = 7.35$ AU and $q = 30$ AU, from top to bottom. A companion plot that presents the regions of each dynamical class can be found in Figure \ref{fig:ae_plane_colors}; the colors of the regions correspond to the marker colors in Figure \ref{fig:ae_plane}.

In Figure \ref{fig:ae_plane}, the current day best-fit $(a,e)$ of each TNO is plotted with a colored marker that corresponds to its dynamical class. The inner centaurs, in red, are found in the giant planet region, with semi-major axes below $a_N = 30$ AU,
and the outer centaurs, in purple, cross Neptune's orbit, with $q < 30$ AU and $a > 30$ AU. Most of the other objects are found near the $q = 30$ AU curve, as it is easier to observe short perihelion TNOs. There are a few exceptions; most notably, a detached TNO in blue with $a = 105$ AU and $q = 50$ AU (\texttt{s17\_good\_0}). 

The population of objects denoted with green markers at low eccentricity constitute the classical belt. These TNOs are dynamically cold (undergo only minimal orbital evolution) as they do not experience strong interactions with Neptune. Their perihelion detachment is evident in the inset plot, which zooms in on the $a \in [30, 60]$, $e \in [0, 0.5]$ region, and demonstrates that the classical belt TNOs have $q > 30$ AU (the solid black curve). In fact, most of these objects have $q = 35 - 37$ AU, as shown in \citet{2011AJ....142..131P}.

Similar to the classical belt population, the detached objects (blue markers) do not interact with Neptune and remain separated from the $q = 30$ AU curve. Defined to be as objects with higher eccentricities, the blue markers are found above the green ones.

The scattering disk objects, marked in yellow, can be found near the $q = 30$ AU curve. These are TNOs with Neptune-driven dynamics, which result in their movement along the $q = 30$ AU curve. The perihelion distance threshold at which objects cease to be affected strongly by Neptune perturbations is often cited to be around $q \approx 35 - 37$ AU \citep{doi:10.1146/annurev.earth.27.1.287, 2007Icar..189..213L}; however, this boundary is actually dependent on semi-major axis \citep{1987AJ.....94.1330D}. Since a TNO's orbital energy scales as $1/a$, at a fixed perihelion distance, larger semi-major axis objects are more strongly affected by energy kicks from Neptune and thus experience greater orbital evolution. 

In the inset, it is possible to note objects marked with dark gray markers; these are the resonant and resonant candidate objects. These TNOs can be found in any region of the phase space, as their location is determined by their semi-major axis alone. For example, in the inset, it is easy to spot the three DES Neptune trojans at the $1$:$1$ resonance at $a = 30$ AU. A more detailed discussion of the resonant TNOs and a plot of the corresponding $a-e$ plane (Figure \ref{fig:res_summary}) are presented in the following section.

\subsection{Resonant Population} \label{sec:res}

\begin{figure}[t!]
\epsscale{1}
  \begin{center}
      \leavevmode
\includegraphics[width=85mm]{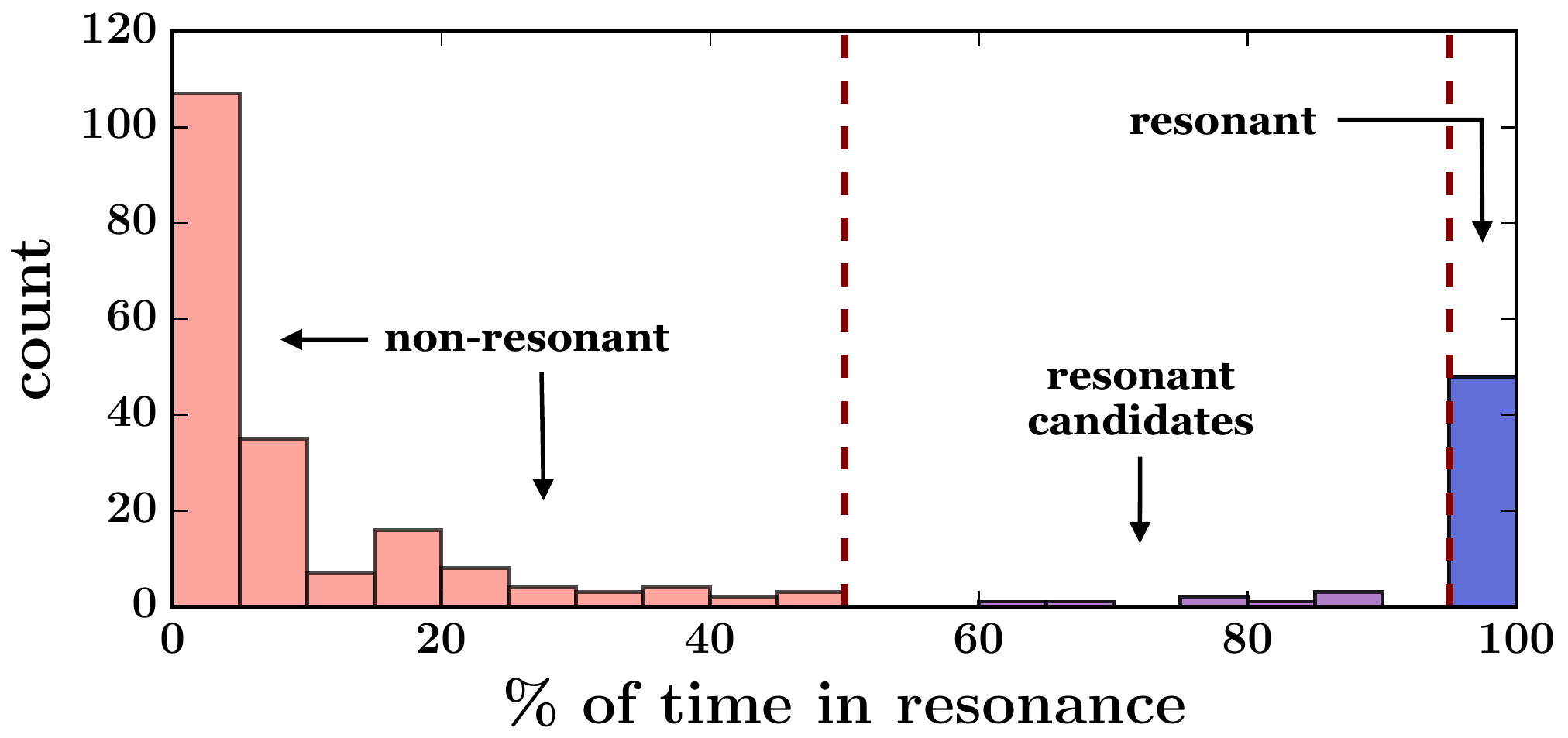}
\caption{A histogram of the percent of time spent in resonance for all of the known DES TNOs. For each object, we compute the percent of time each of its clones spends in resonance and average over all ten clones to find the total resonance percentage. Most objects are securely non-resonant (the 0$\%$ bin). In between the two dashed vertical lines are a few resonant candidates ($50\%$ to $95\%$), and to the right of the line at $95\%$ are the resonant objects: those with clones that are in resonance for the full integration time. }
\label{fig:res_percent}
\end{center}
\end{figure}

The current DES TNO sample contains 48 resonant objects, with an additional 7 resonant candidates, as shown in Figure \ref{fig:res_percent}. In this plot, we present the results of our resonance classification algorithm for the entire DES sample. The histogram displays the percentage of time spent in resonance by each TNO.

To compute this value, we first find the fraction of time each of the ten clones spends in a resonance during the integration time. Sometimes, a clone may visit more than one resonance during the integration; in this case, we take the longest time spent in one resonance. Next, we average over all of the ten clones, and arrive at the percentage of time spent in resonance by each TNO.

The result is shown in the histogram in Figure \ref{fig:res_percent}. Note that there are two peaks of objects - non-resonant TNOs, with 0$\%$ of time spent in resonance, and securely resonant TNOs, with greater than 95$\%$ of time spent in resonance. There are relatively few TNOs in the middle region. This seems to indicate that our resonance-finding algorithm is able to clearly distinguish between the resonant and non-resonant cases, and does not present a large number of semi-resonant objects. 

In reality, objects could indeed be semi-resonant: over long time scales, objects may transition in and out of resonance. The integration times under consideration here, however, are short, and we expect objects to either be resonant or not on these timescales.

In this work, we choose to identify TNOs that are resonant greater than 50$\%$ of the time, but less than 95$\%$ of the time, as resonant candidates. The location of the two thresholds is rather arbitrary, but Figure \ref{fig:res_percent} clearly shows that any reasonable choice will produce quantitatively similar results. The candidate resonant TNOs spend the majority of their time in resonance; as their orbits improve with further observations, it is possible that these TNOs will become securely resonant TNOs.

\begin{figure*}[t!]
\epsscale{1}
  \begin{center}
      \leavevmode
\includegraphics[width=176mm]{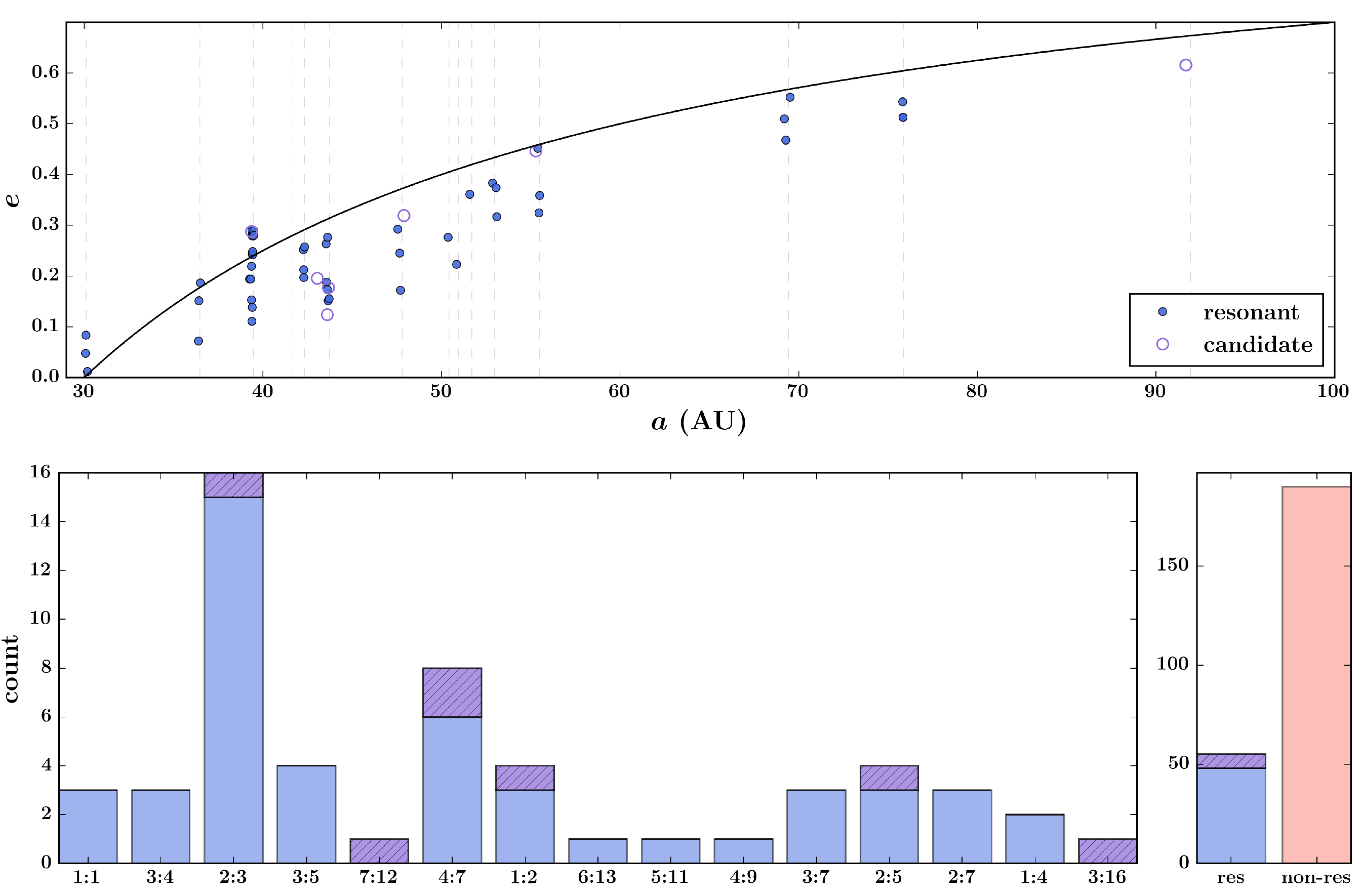}
\caption{The top panel visualizes the resonant TNOs on the $a-e$ plane, while the bottom left panel presents the distribution of resonances for the resonant TNOs. All of the resonant objects are in resonance with Neptune. The most populated resonance is the 2:3 (the Plutinos), and there are a number of higher order resonances, such as the 2:7 or the 6:13. The bottom right panel shows the number of resonant objects as compared to the number of non-resonant TNOs in the DES data. The blue bars represent securely resonant objects, while the purple bars are the resonant candidates. The bottom left bar plot sorts the resonances by period; this allows for easy comparison between the bottom and top panels. For instance, the left three objects in both plots are the three Neptune trojans, and the rightmost TNO in both plots is a $3$:$16$ resonant candidate. The orbital elements of the objects are plotted at the epoch reported in Table \ref{tab:tno_data}.}
\label{fig:res_summary}
\end{center}
\end{figure*}

The DES resonant TNOs populate 15 resonances, ranging from the short period Neptune trojans at the $1$:$1$ resonance to the long period $3$:$16$ TNO candidate. A bar plot of the populated resonances is shown in the bottom panel of Figure \ref{fig:res_summary}. The bottom left panel presents all instances of each $p$:$q$ resonance, sorted by increasing orbital period, from left to right. The blue bars represent the securely resonant objects, and the hatched purple bars show the much smaller population of resonant candidates. The bar plot in the bottom right panel summarizes this data.

The top panel displays the resonant objects on the $a-e$ plane. Note that each resonance is found at a constant semi-major axis as indicated by the dashed vertical lines; as a result, each resonance is reminiscent of beads on a string. Each of these resonances corresponds directly to a bar in the bottom left panel. For example, note the three Neptune trojans on the left in both plots, next the three objects in the $3$:$4$ resonance, and so on.

From this analysis, we see that the resonant TNOs make up a significant portion of the DES dataset, representing about one fifth of the objects. The most populated resonances are the Plutinos, at the $2$:$3$ resonance, but there are a number of higher order resonances in the sample as well.

\section{Discussion} \label{sec:discussion}

In this work, we introduce an updated classification algorithm for the trans-Neptunian region of the Solar System. 
Our classification scheme is fundamentally consistent with the previous classification schemes laid out in \citet{Elliot2005} and \citet{2008ssbn.book...43G}.
Similarly to \citet{Elliot2005}, which used detections from the Deep Ecliptic Survey, we classify a uniformly derived sample of Kuiper Belt objects: all objects were detected so far in the Dark Energy Survey data, many of which are previously undiscovered objects.

Our new resonance-finding tool allows for the automated identification of resonances by using numerical integrations of TNOs, and uses an hierarchical determination of regions where resonance angles librate to identify KBOs in true resonance. Through this method, we classify the current collection of objects detected by the Dark Energy Survey and present a summary of the results. Our classification scheme yields 1 inner centaur, 19 outer centaurs, 21 scattering disk objects, 47 detached TNOs, 48 securely resonant objects, 7 resonant candidates, and 97 classical belt objects.

It is important to note that our classification algorithm is only as good as the certainty of the TNO orbits. Although a poorly constrained orbit can result in a mis-classification in any of the categories, the most sensitive boundary is that for the resonant classification. If the semi-major axis error for a TNO is several AU or more -- greater than a typical resonance width -- then the spread in the initial orbit of the clones will result in overall non-resonant behavior for the TNO. In this situation, the object may be classified as scattering or as a classical belt/detached TNO, depending on its perihelion distance. In reality, however, the TNO could actually be in a resonance, but the wide range of possible semi-major axes $a$ (due to large uncertainties) prevents us from making a secure classification. 

On the other hand, classifying an object as a securely resonant TNO is an indication that it has a well-defined orbit with small errors, and further observations of the object are unlikely to change the classification. That is, general improvement of the orbit certainties for the TNOs could potentially increase the number of objects in the resonance class, and decrease the number of objects in the other classes.

We expect the coming years to witness a substantial increase in the numbers of TNOs detected by DES as the remaining data is analyzed (e.g., \citealt{2019arXiv190901478B}). Once the additional objects are classified and combined with the current dataset, we plan to conduct a suite of population-wide analyses of the TNOs. In combination with the DES survey simulator, such future work will reveal the structure of this distant region and allow for the testing of formation hypotheses of the outer Solar System. 

\acknowledgements
This material is based upon work supported by the National Aeronautics and Space
Administration under Grant No. NNX17AF21G issued through the SSO Planetary Astronomy Program. J.C.B, L.M., and S.J.H are supported by the NSF Graduate Research Fellowship Grant No. DGE 1256260.

This work included the observations obtained at the MDM Observatory, operated by Dartmouth College, Columbia University, Ohio State University, Ohio University, and the University of Michigan.

Funding for the DES Projects has been provided by the U.S. Department of Energy, the U.S. National Science Foundation, the Ministry of Science and Education of Spain, 
the Science and Technology Facilities Council of the United Kingdom, the Higher Education Funding Council for England, the National Center for Supercomputing 
Applications at the University of Illinois at Urbana-Champaign, the Kavli Institute of Cosmological Physics at the University of Chicago, 
the Center for Cosmology and Astro-Particle Physics at the Ohio State University,
the Mitchell Institute for Fundamental Physics and Astronomy at Texas A\&M University, Financiadora de Estudos e Projetos, 
Funda{\c c}{\~a}o Carlos Chagas Filho de Amparo {\`a} Pesquisa do Estado do Rio de Janeiro, Conselho Nacional de Desenvolvimento Cient{\'i}fico e Tecnol{\'o}gico and 
the Minist{\'e}rio da Ci{\^e}ncia, Tecnologia e Inova{\c c}{\~a}o, the Deutsche Forschungsgemeinschaft and the Collaborating Institutions in the Dark Energy Survey. 

The Collaborating Institutions are Argonne National Laboratory, the University of California at Santa Cruz, the University of Cambridge, Centro de Investigaciones Energ{\'e}ticas, 
Medioambientales y Tecnol{\'o}gicas-Madrid, the University of Chicago, University College London, the DES-Brazil Consortium, the University of Edinburgh, 
the Eidgen{\"o}ssische Technische Hochschule (ETH) Z{\"u}rich, 
Fermi National Accelerator Laboratory, the University of Illinois at Urbana-Champaign, the Institut de Ci{\`e}ncies de l'Espai (IEEC/CSIC), 
the Institut de F{\'i}sica d'Altes Energies, Lawrence Berkeley National Laboratory, the Ludwig-Maximilians Universit{\"a}t M{\"u}nchen and the associated Excellence Cluster Universe, 
the University of Michigan, the National Optical Astronomy Observatory, the University of Nottingham, The Ohio State University, the University of Pennsylvania, the University of Portsmouth, 
SLAC National Accelerator Laboratory, Stanford University, the University of Sussex, Texas A\&M University, and the OzDES Membership Consortium.

Based in part on observations at Cerro Tololo Inter-American Observatory, National Optical Astronomy Observatory, which is operated by the Association of 
Universities for Research in Astronomy (AURA) under a cooperative agreement with the National Science Foundation.

The DES data management system is supported by the National Science Foundation under Grant Numbers AST-1138766 and AST-1536171.
The DES participants from Spanish institutions are partially supported by MINECO under grants AYA2015-71825, ESP2015-66861, FPA2015-68048, SEV-2016-0588, SEV-2016-0597, and MDM-2015-0509, 
some of which include ERDF funds from the European Union. IFAE is partially funded by the CERCA program of the Generalitat de Catalunya.
Research leading to these results has received funding from the European Research
Council under the European Union's Seventh Framework Program (FP7/2007-2013) including ERC grant agreements 240672, 291329, and 306478.
We  acknowledge support from the Brazilian Instituto Nacional de Ci\^encia
e Tecnologia (INCT) e-Universe (CNPq grant 465376/2014-2).

This manuscript has been authored by Fermi Research Alliance, LLC under Contract No. DE-AC02-07CH11359 with the U.S. Department of Energy, Office of Science, Office of High Energy Physics. The United States Government retains and the publisher, by accepting the article for publication, acknowledges that the United States Government retains a non-exclusive, paid-up, irrevocable, world-wide license to publish or reproduce the published form of this manuscript, or allow others to do so, for United States Government purposes.

\bigskip
\bigskip




\section*{Supplementary material}

**Table 1.** Barycentric orbital elements for the set of TNOs detected by DES and considered in this work. Some data is obtained from follow up observations, which improves the classification. Numbers are reported to representative errors. 'Res' denotes the specific resonance in which an object lives, if applicable. Solutions are reported at the epoch given in the final column of the table. Objects are ordered by current semi-major axis (in AU) and identified by the MPC identifier (if available) or by their DES internal identifier if an MPC designation does not exist. Angles $(i, \omega, \Omega, M)$ are given in degrees.

| TNO ID | Class | Res. | $a_b$ (AU) | $e_b$ | $i_b$ (deg) | $\omega_b$ (deg) | $\Omega_b$ (deg) | $M_b$ (deg) | Epoch (JD) |
|---|---|---|---|---|---|---|---|---|---|
| 2003 QC112 | inner centaur | - | 20.5004 ± 0.0009 | 0.27616 ± 3 × 10⁻⁵ | 18.245 ± 0.0001 | 22.194 ± 0.009 | 158.6556 ± 0.0003 | 191.708 ± 0.009 | 2456578.73 |
| 2014 UU240 | resonant | 1:1 | 30.056 ± 0.001 | 0.048 ± 0.003 | 35.747 ± 0.004 | 74 ± 3 | 81.993 ± 0.004 | 236 ± 3 | 2456959.83 |
| 2013 VX30 | resonant | 1:1 | 30.0871 ± 0.0006 | 0.08374 ± 2 × 10⁻⁵ | 31.25873 ± 7 × 10⁻⁵ | 215.49 ± 0.02 | 192.53852 ± 8 × 10⁻⁵ | 347.84 ± 0.02 | 2456567.79 |
| (530664) 2011 SO277 | resonant | 1:1 | 30.1614 ± 0.0005 | 0.01185 ± 8 × 10⁻⁵ | 9.6386 ± 0.0002 | 117.7 ± 0.3 | 113.5271 ± 0.0009 | 148.1 ± 0.5 | 2456545.88 |
| (309239) 2007 RW10 | outer centaur | - | 30.236 ± 0.002 | 0.30055 ± 6 × 10⁻⁵ | 36.1011 ± 0.0001 | 96.095 ± 0.005 | 187.03731 ± 7 × 10⁻⁵ | 61.472 ± 0.005 | 2456547.85 |
| 2013 RD109 | outer centaur | - | 32.378 ± 0.001 | 0.08106 ± 2 × 10⁻⁵ | 11.1194 ± 0.0001 | 332.64 ± 0.05 | 16.6919 ± 0.0001 | 11.31 ± 0.05 | 2456537.77 |
| 2014 UC225 | scattering | - | 34.7 +0.4 −0.1 | 0.1 ± 0.04 | 4.942 ± 0.02 | 221 ± 100 | 139.7 ± 0.2 | 17 ± 90 | 2456951.73 |
| 2013 RH109 | resonant | 3:4 | 36.38 ± 0.005 | 0.0725 ± 0.0002 | 14.8214 ± 0.0003 | 288.5 ± 0.3 | 204.394 ± 0.001 | 196.8 ± 0.3 | 2456543.59 |
| 2013 RQ109 | resonant | 3:4 | 36.404 ± 0.003 | 0.1512 ± 6 × 10⁻⁵ | 14.5378 ± 0.003 | 335.91 ± 0.03 | 70.5182 ± 0.0006 | 339.72 ± 0.02 | 2456547.89 |
| 2014 TM95 | resonant | 3:4 | 36.487 ± 0.002 | 0.18664 ± 6 × 10⁻⁵ | 17.5579 ± 0.0001 | 263.3 ± 0.01 | 161.1436 ± 0.0004 | 328.744 ± 0.008 | 2456569.69 |
| 2013 SH102 | classical belt | - | 37.902 ± 0.003 | 0.0528 ± 0.0002 | 19.3819 ± 0.0003 | 83.07 ± 0.07 | 180.22736 ± 2 × 10⁻⁵ | 93.4 ± 0.1 | 2456565.66 |
| 2013 RG109 | classical belt | - | 38.241 ± 0.003 | 0.0904 ± 0.0002 | 22.7219 ± 0.0002 | 59.72 ± 0.04 | 22.6841 ± 0.0001 | 298.16 ± 0.02 | 2456537.86 |
| 2014 RH70 | classical belt | - | 38.251 ± 0.008 | 0.1224 ± 0.0002 | 27.60542 ± 5 × 10⁻⁵ | 244.72 ± 0.06 | 8.0072 ± 0.0004 | 39.53 ± 0.04 | 2456904.6 |
| s200_good_333 | outer centaur | - | 38.71 ± 0.01 | 0.3508 ± 0.0008 | 17.2877 ± 0.0006 | 35.9 ± 0.1 | 183.4814 ± 0.0002 | 100.2 ± 0.1 | 2456543.67 |
| (120348) 2004 TY364 | classical belt | - | 38.86 ± 0.02 | 0.068 ± 0.004 | 24.838 ± 0.001 | 358 ± 0.6 | 140.5 ± 0.006 | 265.8 ± 0.7 | 2456904.85 |
| s13_good_5 | classical belt | - | 39 ± 0.1 | 0.13 ± 0.02 | 38.39 ± 0.01 | 300 ± 6 | 169.46 ± 0.007 | 310 ± 3 | 2456958.85 |
| 2003 QB91 | resonant | 2:3 | 39.241 ± 0.002 | 0.1942 ± 0.0002 | 6.4955 ± 0.0002 | 80.6 ± 0.07 | 136.788 ± 0.003 | 142.4 ± 0.07 | 2456578.71 |
| (534315) 2014 SK349 | resonant candidate | 2:3 | 39.3 ± 0.2 | 0.288 ± 0.003 | 9.41 ± 0.03 | 313 ± 4 | 59.8 ± 0.1 | 6 ± 2 | 2456932.79 |
| 2014 WC536 | resonant | 2:3 | 39.32 ± 0.03 | 0.194 ± 0.002 | 22.435 ± 0.002 | 259.8 ± 0.7 | 88.493 ± 0.004 | 34.2 ± 0.4 | 2456328.59 |
| 2013 SO102 | resonant | 2:3 | 39.353 ± 0.002 | 0.21965 ± 4 × 10⁻⁵ | 9.8516 ± 9.00E-05 | 257.86 ± 0.02 | 145.2264 ± 0.0008 | 346.88 ± 0.01 | 2456564.83 |
| 2013 SP102 | resonant | 2:3 | 39.357 ± 0.001 | 0.15307 ± 8 × 10⁻⁵ | 11.5999 ± 0.0002 | 75.65 ± 0.08 | 146.688 ± 0.001 | 158.75 ± 0.08 | 2456564.84 |
| (469372) 2001 QF298 | resonant | 2:3 | 39.377 ± 0.002 | 0.11115 ± 9.00E-05 | 22.3519 ± 0.0003 | 42.2 ± 0.1 | 164.18428 ± 4 × 10⁻⁵ | 149.8 ± 0.1 | 2456537.84 |

**Table 1.** continued from previous page

| TNO ID | Class | Res. | $a_b$ (AU) | $e_b$ | $i_b$ (deg) | $\omega_b$ (deg) | $\Omega_b$ (deg) | $M_b$ (deg) | Epoch (JD) |
|---|---|---|---|---|---|---|---|---|---|
| 2012 WF37 | resonant | 2:3 | 39.38 ± 0.003 | 0.29036 ± 4×10⁻⁵ | 19.01377 ± 8×10⁻⁵ | 275.142 ± 0.009 | 173.6208 ± 0.0002 | 328.669 ± 0.004 | 2456247.62 |
| s302_good_485 | resonant | 2:3 | 39.384 ± 0.003 | 0.28046 ± 6×10⁻⁵ | 10.36557 ± 4×10⁻⁵ | 285.48 ± 0.01 | 129.4896 ± 0.0009 | 342.594 ± 0.007 | 2456569.7 |
| s302_good_198 | resonant | 2:3 | 39.385 ± 0.003 | 0.28046 ± 6×10⁻⁵ | 10.36557 ± 4×10⁻⁵ | 285.48 ± 0.01 | 129.4897 ± 0.0009 | 342.594 ± 0.007 | 2456569.7 |
| 2012 TD324 | resonant | 2:3 | 39.39 ± 0.002 | 0.1386 ± 0.0001 | 9.57606 ± 8×10⁻⁵ | 191 ± 0.05 | 114.616 ± 0.001 | 49.44 ± 0.03 | 2456544.82 |
| 2013 TY171 | resonant | 2:3 | 39.402 ± 0.002 | 0.24306 ± 9.00E-05 | 24.9545 ± 0.0002 | 247.63 ± 0.01 | 58.8074 ± 0.0003 | 47.476 ± 0.003 | 2456569.72 |
| (504555) 2008 SO266 | resonant | 2:3 | 39.407 ± 0.002 | 0.24255 ± 6×10⁻⁵ | 18.7979 ± 0.0004 | 172.76 ± 0.01 | 158.7544 ± 0.0004 | 34.401 ± 0.005 | 2456569.77 |
| 2014 WD536 | resonant | 2:3 | 39.408 ± 0.003 | 0.24871 ± 7×10⁻⁵ | 16.6069 ± 0.0001 | 329.35 ± 0.03 | 68.2738 ± 0.0005 | 346.26 ± 0.02 | 2456545.8 |
| 2013 RC109 | resonant | 2:3 | 39.419 ± 0.003 | 0.2791 ± 0.0001 | 43.5138 ± 0.0004 | 318.85 ± 0.04 | 32.8166 ± 0.0002 | 14.01 ± 0.02 | 2456544.84 |
| 2013 TA172 | classical belt | - | 39.424 ± 0.004 | 0.18365 ± 9.00E-05 | 14.5476 ± 0.0002 | 237.4 ± 0.002 | 173.6434 ± 0.0002 | 329.27 ± 0.01 | 2456578.64 |
| (534315) 2014 SK349 | resonant | 2:3 | 39.471 ± 0.002 | 0.2897 ± 0.0001 | 9.395 ± 0.0004 | 315.9 ± 0.4 | 59.898 ± 0.0001 | 3 ± 0.2 | 2456569.69 |
| 2010 SB41 | resonant | 2:3 | 39.48 ± 0.002 | 0.28028 ± 4×10⁻⁵ | 5.22363 ± 7×10⁻⁵ | 248.8 ± 0.02 | 139.566 ± 0.001 | 352.74 ± 0.01 | 2456537.86 |
| s121_good_1 | classical belt | - | 39.7 +0.2 -0.1 | 0.05 ± 0.004 | 54.78 ± 0.06 | 256 ± 40 | 206.506 ± -0.009 | 359 +- 300 | 2457014.83 |
| 2014 XZ40 | classical belt | - | 39.794 +0.007 -0.005 | 0.061 ± 0.003 | 44.56 ± 0.02 | 257 ± 5 | 146.807 ± 0.008 | 28 ± 5 | 2456992.8 |
| (505412) 2013 QO95 | classical belt | - | 39.9679 ± 0.0003 | 0.03267 ± 10⁻⁵ | 20.6027 ± 0.0003 | 316.1 ± 0.2 | 83.1093 ± 0.0007 | 349.5 ± 0.1 | 2456534.7 |
| 2013 RL109 | scattering | - | 40.185 ± 0.007 | 0.2015 ± 0.0003 | 14.1841 ± 0.0002 | 69.56 ± 0.02 | 193.5998 ± 0.0006 | 51.46 ± 0.01 | 2456545.56 |
| 2013 RB109 | classical belt | - | 40.208 ± 0.002 | 0.1081 ± 0.0001 | 23.1512 ± 0.0001 | 223.94 ± 0.08 | 175.85643 ± 3×10⁻⁵ | 329.3 ± 0.06 | 2456537.77 |
| s301a_good_186 | classical belt | - | 40.21 +0.01 -0.007 | 0.12 ± 0.008 | 1.7616 ± 0.0003 | 110 ± 20 | 54.63 ± 0.01 | 202 ± 10 | 2456604.64 |
| s301_good_1175 | classical belt | - | 40.25 ± 0.01 | 0.104 ± 0.002 | 24.198 ± 0.0003 | 211.6 ± 0.6 | 32.681 ± 0.002 | 112.3 ± 0.7 | 2456545.83 |
| s240_good_3 | classical belt | - | 40.3 +0.1 -0.004 | 0.08 ± 0.01 | 22.55 ± 0.002 | 238 ± 20 | 43.49 ± 0.04 | 22 ± 20 | 2456877.66 |
| s200_good_407 | classical belt | - | 40.33 +0.02 -0.05 | 0.06 ± 0.01 | 14.6 ± 0.03 | 101 ± 20 | 268.65 ± -0.04 | 323 ± 20 | 2456538.7 |
| s200_good_743 | classical belt | - | 40.35 ± 0.003 | 0.07129 ± 5×10⁻⁵ | 18.2052 ± 0.0003 | 154.5 ± 0.4 | 188.986 ± 0.0004 | 2.6 ± 0.3 | 2456546.74 |
| 2014 TF86 | classical belt | - | 40.421 ± 0.006 | 0.0782 ± 0.0002 | 32.0479 ± 0.0002 | 106.2 ± 0.2 | 65.0835 ± 0.0006 | 142.1 ± 0.2 | 2456931.58 |
| s241_good_4 | classical belt | - | 40.5 ± 0.05 | 0.109 ± 0.006 | 38.464 ± 0.004 | 307.3 ± 0.9 | 87.574 ± 0.007 | 295.9 ± 0.2 | 2456903.64 |
| 2013 QP95 | classical belt | - | 40.6434 ± 0.0009 | 0.16937 ± 5×10⁻⁵ | 25.4409 ± 0.0001 | 18.79 ± 0.01 | 71.3968 ± 0.0002 | 312.537 ± 0.005 | 2456534.7 |
| s200_good_658 | outer centaur | - | 40.83 ± 0.02 | 0.2745 ± 0.0005 | 27.9843 ± 0.0008 | 193.4 ± 0.3 | 175.65086 ± 8×10⁻⁵ | 339.32 ± 0.07 | 2456545.56 |
| 2015 TK363 | classical belt | - | 40.888 ± 0.003 | 0.0664 ± 0.0002 | 14.7881 ± 0.0002 | 174.36 ± 0.06 | 142.995 ± 0.001 | 60.21 ± 0.05 | 2456654.61 |
| s242_good_7 | classical belt | - | 40.971 +0.009 -0.008 | 0.043 ± 0.0005 | 32.369 ± 0.002 | 255 ± 1 | 50.97 ± 0.001 | 17 ± 1 | 2456559.57 |

**Table 1.** continued from previous page

| TNO ID | Class | Res. | $a_b$ (AU) | $e_b$ | $i_b$ (deg) | $\omega_b$ (deg) | $\Omega_b$ (deg) | $M_b$ (deg) | Epoch (JD) |
|---|---|---|---|---|---|---|---|---|---|
| (145452) 2005 RN43 | classical belt | - | $41.512$ $^{+0.008}_{-0.007}$ | $0.0225$ $\pm$ $0.0003$ | $19.2711$ | $172 \pm 2$ | $186.9928$ $\pm$ $0.0002$ | $338 \pm 2$ | 2456543.64 |
| 2014 UF241 | classical belt | - | $41.516$ $\pm$ $0.006$ | $0.1591$ $\pm$ $0.0003$ | $26.75$ $\pm$ $0.0009$ | $18 \pm 0.3$ | $191.5667$ $\pm$ $0.0004$ | $169.5$ $\pm$ $0.3$ | 2456951.76 |
| 2014 PS70 | classical belt | - | $41.7$ $^{+0.3}_{-0.2}$ | $0.09$ $\pm$ $0.02$ | $15.1864$ $\pm$ $0.003$ | $308 \pm 7$ | $347.08$ $\pm$ $-0.03$ | $51 \pm 5$ | 2457661.59 |
| 2014 VW37 | classical belt | - | $42.083$ $\pm$ $0.002$ | $0.13267$ $\pm$ $7 \times$ $10^{-5}$ | $48.7849$ $\pm$ $0.0002$ | $255.5$ $\pm$ $0.06$ | $122.7768$ $\pm$ $0.0003$ | $16.64$ $\pm$ $0.05$ | 2456973.85 |
| (503883) 2001 QF331 | resonant | 3:5 | $42.251$ $\pm$ $0.009$ | $0.2524$ $\pm$ $0.0003$ | $2.673$ $\pm$ $0.0005$ | $249.4$ $\pm$ $0.1$ | $156.785$ $\pm$ $0.008$ | $339.68$ $\pm$ $0.05$ | 2456544.7 |
| 2012 TC324 | resonant | 3:5 | $42.27$ $\pm$ $0.002$ | $0.19703$ $\pm$ $3 \times$ $10^{-5}$ | $9.631$ $\pm$ $4 \times$ $10^{-5}$ | $259.27$ $\pm$ $0.02$ | $131.8323$ $\pm$ $0.0009$ | $357.04$ $\pm$ $0.01$ | 2456569.7 |
| 2013 UT22 | resonant | 3:5 | $42.275$ $\pm$ $0.002$ | $0.21235$ $\pm$ $8 \times$ $10^{-5}$ | $29.3413$ $\pm$ $0.0001$ | $138.47$ $\pm$ $0.01$ | $194.2379$ $\pm$ $0.0002$ | $44.05$ $\pm$ $0.005$ | 2456268.65 |
| 2013 TH172 | resonant | 3:5 | $42.313$ $\pm$ $0.003$ | $0.25773$ $\pm$ $7 \times$ $10^{-5}$ | $11.5541$ $\pm$ $0.0002$ | $287.88$ $\pm$ $0.02$ | $42.6569$ $\pm$ $0.0004$ | $24.02$ $\pm$ $0.01$ | 2456568.61 |
| 2014 UD241 | classical belt | - | $42.331$ $\pm$ $0.003$ | $0.0498$ $\pm$ $0.0002$ | $12.4909$ $\pm$ $0.0003$ | $68.64$ $\pm$ $0.09$ | $40.6426$ $\pm$ $0.0009$ | $264.13$ $\pm$ $0.08$ | 2456619.64 |
| 2014 UE241 | classical belt | - | $42.476$ $\pm$ $0.005$ | $0.1234$ $\pm$ $0.0002$ | $7.5128$ $\pm$ $0.0003$ | $78.37$ $\pm$ $0.09$ | $60.419$ $\pm$ $0.002$ | $242.9$ $\pm$ $0.1$ | 2456569.66 |
| 2013 TD172 | scattering | - | $42.5$ $\pm$ $0.1$ | $0.163$ $\pm$ $0.007$ | $11.003$ $\pm$ $0.001$ | $85.4$ $\pm$ $0.3$ | $6.0387$ $\pm$ $0.0004$ | $294.5$ $\pm$ $0.3$ | 2456578.59 |
| 2013 RN109 | classical belt | - | $42.536$ $\pm$ $0.003$ | $0.1556$ $\pm$ $9.00E$-$05$ | $32.3057$ $\pm$ $0.0002$ | $17.89$ $\pm$ $0.04$ | $20.9369$ $\pm$ $0.0001$ | $336.06$ $\pm$ $0.03$ | 2456545.84 |
| 2003 SQ317 | classical belt | - | $42.659$ $\pm$ $0.002$ | $0.08003$ $\pm$ $3 \times$ $10^{-5}$ | $28.5671$ $\pm$ $0.0002$ | $193.1$ $\pm$ $0.09$ | $176.30698$ $\pm$ $5 \times$ $10^{-5}$ | $0.65$ $\pm$ $0.08$ | 2456537.85 |
| 2013 UQ15 | classical belt | - | $42.77$ $\pm$ $0.002$ | $0.113$ $\pm$ $0.0001$ | $27.3432$ $\pm$ $0.0005$ | $15.8$ $\pm$ $0.2$ | $189.1313$ $\pm$ $0.0003$ | $169.8$ $\pm$ $0.2$ | 2456932.79 |
| 2014 SN350 | classical belt | - | $42.82$ $\pm$ $0.003$ | $0.18878$ $\pm$ $9.00E$-$05$ | $32.0584$ $\pm$ $0.0002$ | $258.54$ $\pm$ $0.04$ | $171.6813$ $\pm$ $0.0002$ | $333.54$ $\pm$ $0.02$ | 2456925.78 |
| s11_good_20 | classical belt | - | $42.82$ $\pm$ $0.004$ | $0.1711$ $\pm$ $0.0001$ | $22.7112$ $\pm$ $0.0001$ | $245.05$ $\pm$ $0.08$ | $107.6522$ $\pm$ $0.0008$ | $17.65$ $\pm$ $0.05$ | 2456888.86 |
| 2013 RF109 | classical belt | - | $42.864$ $\pm$ $0.002$ | $0.06475$ $\pm$ $6 \times$ $10^{-5}$ | $9.7153$ $\pm$ $0.0002$ | $243.6$ $\pm$ $0.1$ | $144.8564$ $\pm$ $0.0001$ | $334.32$ $\pm$ $0.09$ | 2456537.84 |
| s13_good_9 | classical belt | - | $42.899$ $\pm$ $0.002$ | $0.16324$ $\pm$ $7 \times$ $10^{-5}$ | $32.8648$ $\pm$ $0.0002$ | $251.8$ $\pm$ $0.08$ | $140.8101$ $\pm$ $0.0005$ | $7.87$ $\pm$ $0.06$ | 2456920.84 |
| 2001 QO297 | classical belt | - | $42.933$ $\pm$ $0.002$ | $0.0365$ $\pm$ $0.0002$ | $1.1363$ $\pm$ $0.0002$ | $316.5$ $\pm$ $0.06$ | $143.319$ $\pm$ $0.008$ | $267.33$ $\pm$ $0.06$ | 2456543.69 |
| 2013 RP98 | classical belt | - | $42.934$ $\pm$ $0.004$ | $0.1318$ $\pm$ $0.0002$ | $13.288$ $\pm$ $0.0001$ | $177.85$ $\pm$ $0.06$ | $216.3248$ $\pm$ $0.0009$ | $306.75$ $\pm$ $0.03$ | 2456538.7 |
| (160256) 2002 PD149 | classical belt | - | $42.954$ $\pm$ $0.004$ | $0.0615$ $\pm$ $0.0003$ | $4.9073$ $\pm$ $0.0001$ | $39 \pm 0.2$ | $103.551$ $\pm$ $0.003$ | $221.4$ $\pm$ $0.2$ | 2456543.69 |
| (160256) 2002 PD149 | classical belt | - | $43 \pm 1$ | $0.06$ $\pm$ $0.06$ | $4.909$ $\pm$ $0.007$ | $38 \pm 100$ | $103.647$ $\pm$ $0.008$ | $222$ $\pm$ $100$ | 2456543.69 |
| 2003 QZ111 | classical belt | - | $43.008$ $\pm$ $0.002$ | $0.06081$ $\pm$ $4 \times$ $10^{-5}$ | $2.6596$ $\pm$ $0.0002$ | $16.6$ $\pm$ $0.2$ | $326.046$ $\pm$ $0.002$ | $12.1$ $\pm$ $0.1$ | 2456565.63 |
| 2013 TB172 | resonant candidate | 7:12 | $43.033$ $\pm$ $0.003$ | $0.19565$ $\pm$ $5 \times$ $10^{-5}$ | $11.6245$ $\pm$ $0.0003$ | $327.52$ $\pm$ $0.08$ | $35.2157$ $\pm$ $0.0006$ | $4.76$ $\pm$ $0.05$ | 2456578.71 |
| 2013 SG102 | classical belt | - | $43.072$ $\pm$ $-0.001$ | $0.0074$ $\pm$ $0.0001$ | $7.9861$ $\pm$ $0.0003$ | $201 \pm 2$ | $200.565$ $\pm$ $0.002$ | $306.67$ $\pm$ $-0.06$ | 2456565.62 |
| 2003 QM91 | classical belt | - | $43.1259$ $\pm$ $0.0008$ | $0.04383$ $\pm$ $3 \times$ $10^{-5}$ | $3.0432$ $\pm$ $0.0005$ | $8.7 \pm 0.2$ | $8.064$ $\pm$ $-0.0002$ | $350 \pm 0.2$ | 2456543.69 |
| (385201) 1999 RN215 | classical belt | - | $43.1648$ $\pm$ $0.0007$ | $0.0733$ $\pm$ $0.0003$ | $12.4099$ $\pm$ $0.0005$ | $102.4$ $\pm$ $0.3$ | $140.65$ $\pm$ $0.002$ | $137.8$ $\pm$ $0.4$ | 2456931.84 |

**Table 1**. continued from previous page

| TNO ID | Class | Res. | $a_b$ (AU) | $e_b$ | $i_b$ (deg) | $\omega_b$ (deg) | $\Omega_b$ (deg) | $M_b$ (deg) | Epoch (JD) |
|---|---|---|---|---|---|---|---|---|---|
| 2013 RJ109 | classical belt | - | 43.32 ± 0.005 | 0.1318 ± 0.0004 | 20.2991 ± 0.06 | 253.57 ± 0.0002 | 182.4743 ± 0.02 | 286.25 ± 0.02 | 2456543.66 |
| (471954) 2013 RM98 | classical belt | - | 43.329 ± 0.005 | 0.1313 ± 0.0003 | 28.0859 ± 0.0003 | 252.68 ± 0.06 | 352.43819 ± 9.00E-05 | 94.47 ± 0.08 | 2456543.67 |
| s301_good_1073 | classical belt | - | 43.3342 ± 0.0002 | 0.0592 ± 0.0003 | 4.2203 | 90 ± 1 | 98.73 ± 0.01 | 184 ± 1 | 2456578.68 |
| s200_good_25 | classical belt | - | 43.4 ± 0.2 | 0.15 ± 0.01 | 6.236 | 117 ± 4 | 283.68 ± 0.02 | 317 ± 2 | 2457614.77 |
| s301_good_2580 | classical belt | - | 43.428 ± 0.001 | 0.0379 ± 0.0002 | 4.1259 ± 0.0005 | 174.3 ± 0.3 | 44.304 ± 0.003 | 141.7 ± 0.4 | 2456546.8 |
| 2013 SK102 | classical belt | - | 43.482 ± 0.003 | 0.1837 ± 0.0002 | 7.4184 ± 0.0005 | 151.4 ± 0.1 | 66.342 ± 0.003 | 134 ± 0.1 | 2456563.62 |
| 2014 QU495 | resonant | 4:7 | 43.527 ± 0.005 | 0.2635 ± 0.0002 | 21.2897 ± 0.03 | 170.21 ± 0.0003 | 177.5986 ± 0.02 | 22.17 ± 0.02 | 2456887.86 |
| 2013 VZ31 | classical belt | - | 43.53 ± 0.005 | 0.1171 ± 0.0003 | 2.6737 ± 0.0001 | 138.8 ± 0.2 | 74.856 ± 0.008 | 139.2 ± 0.2 | 2456604.66 |
| s302_good_82 | classical belt | - | 43.54 ± 0.001 | 0.09362 ± $7 \times 10^{-5}$ | 10.0144 ± 0.0001 | 281.9 ± 0.2 | 117.109 ± 0.002 | 347.9 ± 0.2 | 2456568.79 |
| s118_good_10 | classical belt | - | 43.552 ± 0.003 | 0.0719 ± 0.0007 | 35.806 ± 0.005 | 47.1 ± 0.2 | 72.238 ± 0.001 | 261.1 ± 0.1 | 2457017.64 |
| 2014 TL95 | resonant | 4:7 | 43.556 ± 0.004 | 0.1881 ± 0.0001 | 10.5949 ± 0.0001 | 307.06 ± 0.07 | 100.919 ± 0.002 | 344.64 ± 0.04 | 2456931.84 |
| (119956) 2002 PA149 | resonant | 4:7 | 43.587 ± 0.004 | 0.174 ± 0.004 | 4.04955 ± $8 \times 10^{-5}$ | 153.1 ± 0.02 | 105.579 ± 0.04 | 81.91 ± 0.04 | 2456537.78 |
| s200_good_481 | resonant candidate | 4:7 | 43.6 $^{+0.03}_{-0.05}$ | 0.124 ± 0.004 | 11.63 ± 0.02 | 234 ± 10 | 297.65 ± -0.02 | 172 ± 10 | 2456548.67 |
| 2013 SJ102 | resonant | 4:7 | 43.617 ± 0.004 | 0.27648 ± $8 \times 10^{-5}$ | 7.34556 ± $2 \times 10^{-5}$ | 315.52 ± 0.01 | 93.723 ± 0.001 | 334.381 ± 0.005 | 2456563.62 |
| 2013 RE109 | resonant | 4:7 | 43.649 ± 0.002 | 0.1519 ± $5 \times 10^{-5}$ | 5.41702 ± $10^{-5}$ | 165.33 ± 0.02 | 112.863 ± 0.002 | 72.9 ± 0.01 | 2456537.79 |
| s301_good_1198 | resonant candidate | 4:7 | 43.7 ± 0.1 | 0.177 ± 0.005 | 2.3817 ± 0.0002 | 250 ± 2 | 80.72 ± 0.01 | 26 ± 1 | 2456546.8 |
| 2001 QE298 | resonant | 4:7 | 43.71 ± 0.01 | 0.1552 ± 0.0001 | 3.6584 ± 0.0004 | 10.7 ± 0.1 | 7.75572 ± $-3 \times 10^{-5}$ | 352 ± 0.1 | 2456593.57 |
| 2013 TF172 | classical belt | - | 43.752 ± 0.004 | 0.0249 ± 0.0002 | 2.8936 ± 0.0002 | 314.2 ± 0.3 | 126.995 ± 0.005 | 284.4 ± 0.2 | 2456578.63 |
| (307616) 2003 QW90 | classical belt | - | 43.765 ± 0.005 | 0.0764 ± 0.0008 | 10.359 ± 0.0007 | 87.23 ± 0.05 | 17.7681 ± 0.0003 | 275.01 ± 0.09 | 2456618.58 |
| 2013 TL172 | classical belt | - | 43.78 ± 0.003 | 0.0611 ± 0.0001 | 1.7911 ± $2 \times 10^{-5}$ | 76.3 ± 0.3 | 95.08 ± 0.01 | 193 ± 0.3 | 2456578.61 |
| s301_good_1491 | classical belt | - | 43.896 ± 0.007 | 0.088 ± 0.0004 | 4.5865 ± 0.0005 | 176.4 ± 0.5 | 35.07 ± 0.003 | 150.2 ± 0.5 | 2456578.61 |
| 2014 VV39 | classical belt | - | 43.938696 ± $10^{-5}$ | 0.0137 ± 0.0001 | 1.6287 ± 0.0001 | 282.2 ± 0.5 | 136.321 ± 0.006 | 310.3 ± 0.2 | 2456546.8 |
| 2001 QQ322 | classical belt | - | 43.991 ± 0.002 | 0.0518 ± 0.0002 | 3.95831 ± $8 \times 10^{-5}$ | 350.6 ± 0.09 | 76.478 ± 0.003 | 297.74 ± 0.05 | 2456544.71 |
| s301_good_1446 | classical belt | - | 44.008 ± -0.007 | 0.0826 ± 0.0004 | 2.9743 ± 0.0001 | 322 ± 2 | 30.279 ± 0.001 | 13 ± 2 | 2456604.67 |
| 2001 QS322 | classical belt | - | 44.02441 ± $5 \times 10^{-5}$ | 0.03869 ± $4 \times 10^{-5}$ | 0.247 ± 0.0004 | 359.4 ± 0.5 | 348.46 ± 0.01 | 15.4 ± 0.5 | 2456565.66 |
| 2013 RO109 | classical belt | - | 44.037 ± 0.001 | 0.03526 ± $6 \times 10^{-5}$ | 1.5237 ± 0.0002 | 328.3 ± 0.3 | 52.095 ± 0.007 | 341.7 ± 0.3 | 2456546.8 |
| s200_good_80 | outer centaur | - | 44.1 ± 0.2 | 0.478 ± 0.003 | 5.0961 ± 0.0008 | 275.9 ± 0.3 | 302.259 ± 0.003 | 75.4 ± 0.7 | 2456540.63 |
| s200_good_750 | classical belt | - | 44.1 ± 0.6 | 0.16 ± 0.05 | 18.29 ± 0.03 | 95 ± 10 | 181.22 ± 0.05 | 41 ± 9 | 2456540.58 |
| s301a_good_324 | classical belt | - | 44.14 ± -0.01 | 0.0113 ± 0.0002 | 1.5949 ± 0.0002 | 313 ± 2 | 63.17 ± 0.01 | 346 ± 2 | 2456578.61 |

**Table 1.** continued from previous page

| TNO ID | Class | Res. | $a_b$ (AU) | $e_b$ | $i_b$ (deg) | $\omega_b$ (deg) | $\Omega_b$ (deg) | $M_b$ (deg) | Epoch (JD) |
|---|---|---|---|---|---|---|---|---|---|
| 2013 WG114 | classical belt | - | 44.151 ± 0.005 | 0.06413 ± 9.00E-05 | 1.4709 | 273.6 ± 0.2 | 70.02 ± 0.01 | 16 ± 0.2 | 2456618.57 |
| s302_good_3 | classical belt | - | 44.251 ± 0.006 | 0.1199 ± 0.0001 | 11.2737 ± 0.0001 | 280.1 ± 0.1 | 114.575 ± 0.002 | 350.29 ± 0.09 | 2456568.79 |
| 2014 OD394 | classical belt | - | 44.365 ± 0.002 | 0.0954 ± 0.0002 | 11.2482 ± 0.0003 | 82.6 ± 0.1 | 130.2 ± 0.001 | 149.3 ± 0.1 | 2456576.73 |
| 2015 RT245 | classical belt | - | 44.384 ± 0.002 | 0.0841 ± 0.0001 | 0.9578 ± 0.0002 | 343.4 ± 0.1 | 330.387 ± 0.007 | 42.18 ± 0.08 | 2456577.62 |
| 2014 VV39 | classical belt | - | 44.45 ± 0.01 | 0.0219 ± 0.0002 | 1.6383 ± 0.0004 | 239 ± 2 | 136.74 ± 0.02 | 352 ± 2 | 2456546.8 |
| s301_good_300 | classical belt | - | 44.63 ± 0.01 | 0.0233 ± 0.0002 | 1.1637 | 320 ± 2 | 49.154 ± 0.007 | 359 ± 400 | 2456563.61 |
| 2013 RP109 | classical belt | - | 44.703 ± 0.002 | 0.10351 ± $4 \times 10^{-5}$ | 2.35056 ± $3 \times 10^{-5}$ | 269.37 ± 0.06 | 105.547 ± 0.005 | 351.16 ± 0.05 | 2456546.8 |
| s240_good_7 | classical belt | - | 44.77 ± 0.03 | 0.1391 ± 0.0005 | 34.3239 ± 0.0002 | 326.2 ± 0.1 | 3.2324 ± 0.0008 | 334.7 ± 0.09 | 2456538.55 |
| 2013 RS109 | classical belt | - | 44.8 ± 0.2 | 0.13 ± 0.0003 | 4.8469 | 305 ± 2 | 353.502 ± -0.004 | 47.9 ± 0.9 | 2456537.76 |
| 2016 SV58 | scattering | - | 44.915 ± 0.005 | 0.2672 ± $6 \times 10^{-5}$ | 13.59559 | 305.7 ± 0.03 | 132.576 ± 0.001 | 333.16 ± 0.02 | 2456953.8 |
| 2001 QO297 | classical belt | - | 45 ± 2 | 0.15 ± 0.08 | 1.138 ± 0.008 | 305 ± 2 | 143.4 ± 0.3 | 292 ± 7 | 2456543.69 |
| 2001 QP297 | classical belt | - | 45.205 ± 0.004 | 0.1206 ± 0.0003 | 1.43081 ± $8 \times 10^{-5}$ | 164.1 ± 0.04 | 111.81 ± 0.01 | 77.22 ± 0.03 | 2456578.61 |
| 2015 PF312 | classical belt | - | 45.2649 ± 0.0006 | 0.09081 ± $2 \times 10^{-5}$ | 17.9941 ± $5 \times 10^{-5}$ | 260.23 ± 0.02 | 160.6353 ± 0.0003 | 337.97 ± 0.02 | 2456247.58 |
| 2013 RX108 | classical belt | - | 45.295 ± 0.003 | 0.0593 ± 0.0002 | 4.8705 ± 0.0001 | 31.76 ± 0.03 | 71.754 ± 0.002 | 277.69 ± 0.04 | 2456537.86 |
| s301_good_946 | classical belt | - | 45.31 ± 0.004 | 0.1718 ± 0.0008 | 16.9 ± 0.001 | 336 ± 1 | 21.2301 ± 0.0008 | 11.1 ± 0.8 | 2456564.73 |
| s200_good_198 | classical belt | - | 45.338 ± 0.005 | 0.1573 ± 0.0002 | 15.1767 ± 0.0003 | 78 ± 0.03 | 184.3703 ± 0.0002 | 73.21 ± 0.03 | 2456546.77 |
| s301_good_1346 | classical belt | - | 45.348 ± 0.002 | 0.08049 ± $4 \times 10^{-5}$ | 1.18649 ± $4 \times 10^{-5}$ | 249.8 ± 0.2 | 109.14 ± 0.01 | 7.6 ± 0.2 | 2456563.61 |
| 2013 TZ171 | classical belt | - | 45.367 ± 0.004 | 0.19211 ± $7 \times 10^{-5}$ | 15.8862 ± 0.0002 | 241.31 ± 0.06 | 163.6584 ± 0.0006 | 349.44 ± 0.04 | 2456569.79 |
| 2013 RY108 | outer centaur | - | 45.53 ± 0.02 | 0.4609 ± 0.0003 | 10.75959 ± $3 \times 10^{-5}$ | 5.171 ± 0.009 | 93.27 ± 0.001 | 321.074 ± 0.008 | 2456545.83 |
| 2014 TB86 | classical belt | - | 45.56 ± 0.002 | 0.17687 ± $4 \times 10^{-5}$ | 19.113 ± 0.0002 | 330.56 ± 0.03 | 50.2367 ± 0.0004 | 353.48 ± 0.02 | 2456545.84 |
| 2014 TB86 | classical belt | - | 45.562 ± 0.002 | 0.17691 ± $3 \times 10^{-5}$ | 19.1129 ± 0.0001 | 330.57 ± 0.02 | 50.2367 ± 0.0003 | 353.47 ± 0.02 | 2456545.84 |
| s301_good_127 | classical belt | - | 45.598 ± 0.003 | 0.14412 ± $6 \times 10^{-5}$ | 6.4625 ± 0.0001 | 269.76 ± 0.06 | 67.99 ± 0.002 | 14.41 ± 0.04 | 2456546.73 |
| 2014 QF442 | classical belt | - | 45.9 ± 0.008 | 0.2071 ± 0.003 | 30.5067 ± $8 \times 10^{-5}$ | 246.4 | 52.9177 ± 0.0009 | 13 ± 0.1 | 2456885.73 |
| s14_good_1 | classical belt | - | 46.066 ± 0.008 | 0.1587 ± 0.0005 | 29.9093 ± 0.0001 | 339.54 ± 0.02 | 131.1873 ± 0.0009 | 309.76 ± 0.03 | 2456916.86 |
| 2015 TJ363 | classical belt | - | 46.132 ± 0.003 | 0.1807 ± 0.0001 | 14.33619 ± 9.00E-05 | 354.15 ± 0.02 | 97.386 ± 0.001 | 309.46 ± 0.01 | 2456569.7 |
| s200_good_615 | scattering | - | 46.312 ± 0.008 | 0.3436 ± 0.0001 | 14.3886 ± 0.0001 | 86.49 ± 0.02 | 202.6127 ± 0.0005 | 16.486 ± 0.009 | 2456543.63 |

**Table 1.** continued from previous page

| TNO ID | Class | Res. | $a_b$ (AU) | $e_b$ | $i_b$ (deg) | $\omega_b$ (deg) | $\Omega_b$ (deg) | $M_b$ (deg) | Epoch (JD) |
|---|---|---|---|---|---|---|---|---|---|
| 2013 TK172 | classical belt | - | 46.459 ± 0.006 | 0.2041 ± 0.0003 | 12.5919 ± 0.0002 | 246.95 ± 0.06 | 71.977 ± 0.001 | 39.01 ± 0.03 | 2456569.67 |
| 2013 TM159 | classical belt | - | 46.468 ± 0.003 | 0.16791 ± 6 × 10⁻⁵ | 9.54264 ± 7 × 10⁻⁵ | 294.31 ± 0.05 | 107.96 ± 0.001 | 347.88 ± 0.03 | 2456568.75 |
| s301d_good_25 | classical belt | - | 46.62 ± 0.01 | 0.1464 ± 0.0007 | 23.726 ± 0.0001 | 95.7 ± 0.2 | 37.028 ± 0.001 | 252.5 ± 0.2 | 2456933.8 |
| 2011 SW281 | classical belt | - | 46.655 ± 0.004 | 0.0949 ± 0.0001 | 4.63342 ± 5 × 10⁻⁵ | 302.61 ± 0.07 | 111.086 ± 0.003 | 323.11 ± 0.05 | 2456569.66 |
| s200_good_540 | classical belt | - | 46.68 ± 0.08 | 0.166 ± 0.005 | 21.6188 ± 0.0004 | 61.5 ± 0.3 | 179.90758 ± 6 × 10⁻⁵ | 77.2 ± 0.7 | 2456548.68 |
| 2014 RJ70 | classical belt | - | 46.88 ± 0.01 | 0.189 ± 0.0002 | 26.47783 ± 2 × 10⁻⁵ | 274.9 ± 0.1 | 30.8855 ± 0.0007 | 358.65 ± 0.07 | 2456912.56 |
| s301_good_160 | classical belt | - | 47 ± 0.3 | 0.2 ± 0.02 | 13.356 ± 0.005 | 357 ± 300 | 43.338 ± 0.001 | 336 ± 7 | 2457639.86 |
| (483002) 2014 QS441 | classical belt | - | 47.0073 ± 0.0008 | 0.08342 ± 8 × 10⁻⁵ | 37.8853 ± 0.0001 | 267.61 ± 0.04 | 185.9271 ± 0.0001 | 306.63 ± 0.02 | 2456594.65 |
| 2013 SS102 | classical belt | - | 47.258 ± 0.004 | 0.1968 ± 0.0001 | 26.3777 ± 0.0002 | 19.65 ± 0.04 | 21.5278 ± 0.0001 | 333.62 ± 0.02 | 2456565.67 |
| s301_good_798 | classical belt | - | 47.49 ± 0.01 | 0.2018 ± 0.0005 | 6.68053 ± 7 × 10⁻⁵ | 20.38 ± 0.07 | 96.327 ± 0.006 | 274.8 ± 0.1 | 2456974.63 |
| (137295) 1999 RB216 | resonant | 1:2 | 47.547 ± 0.003 | 0.29237 ± 3 × 10⁻⁵ | 12.6879 ± 0.0002 | 208.67 ± 0.02 | 175.7239 ± 0.0004 | 359.31 ± 0.01 | 2456545.87 |
| 2012 WE37 | resonant | 1:2 | 47.648 ± 0.005 | 0.24566 ± 6 × 10⁻⁵ | 25.6882 ± 0.0003 | 331.06 ± 0.05 | 59.8663 ± 0.0004 | 0.06 ± 0.03 | 2456247.63 |
| (145452) 2005 RN43 | outer centaur | - | 47.67 ± 0.04 | 0.671 ± 0.0003 | 33.258 ± 0.0005 | 289.7 ± 0.04 | 172.678 ± 0.001 | 314.01 ± 0.02 | 2456575.59 |
| (495189) 2012 VR113 | resonant | 1:2 | 47.692 ± 0.002 | 0.17188 ± 7 × 10⁻⁵ | 19.28378 ± 3 × 10⁻⁵ | 220.45 ± 0.02 | 121.0325 ± 0.0005 | 35.16 ± 0.01 | 2456242.66 |
| 2013 TG172 | resonant candidate | 1:2 | 47.88 ± 0.01 | 0.3196 ± 0.0002 | 4.8011 ± 0.0004 | 339.44 ± 0.05 | 14.472 ± 0.0005 | 6.18 ± 0.02 | 2456568.59 |
| 2013 RR109 | classical belt | - | 48.01 ± 0.03 | 0.066 ± 0.005 | 4.237 ± 0.002 | 90 ± 5 | 225.8 ± 0.02 | 28 ± 5 | 2456546.76 |
| 2013 TE172 | detached | - | 48.255 ± 0.003 | 0.26045 ± 4 × 10⁻⁵ | 29.841 ± 0.0002 | 325.68 ± 0.002 | 50.8138 ± 0.0002 | 2.8 ± 0.01 | 2456569.77 |
| 2016 TY94 | detached | - | 48.855 ± 0.004 | 0.24573 ± 5 × 10⁻⁵ | 25.6687 ± 0.0001 | 279.1 ± 0.07 | 108.7249 ± 0.0005 | 355.81 ± 0.04 | 2456930.75 |
| 2012 WG37 | scattering | - | 49.015 ± 0.009 | 0.3435 ± 0.0002 | 14.3201 ± 0.0002 | 41.17 ± 0.02 | 106.999 ± 0.002 | 288.17 ± 0.04 | 2456250.62 |
| (534073) 2014 QL441 | detached | - | 49.1 ± 0.5 | 0.27 ± 0.02 | 26.27 ± 0.02 | 285 ± 6 | 75.85 ± 0.04 | 14 ± 3 | 2456887.77 |
| 2010 JJ210 | detached | - | 49.28 ± 0.08 | 0.249 ± 0.002 | 7.1674 ± 0.0005 | 101.6 ± 0.7 | 215.93 ± 0.004 | 17.1 ± 0.4 | 2456543.66 |
| s200_good_569 | detached | - | 49.4 ± 0.6 | 0.29 ± 0.02 | 5.617 ± 0.001 | 135 ± 1 | 291.05 ± 0.03 | 316.3 ± 0.1 | 2456545.79 |
| 2016 SP56 | detached | - | 49.64 ± 0.01 | 0.257 ± 0.0002 | 20.0422 ± 0.0003 | 30.74 ± 0.04 | 75.2708 ± 0.0008 | 313.33 ± 0.02 | 2456888.89 |
| s119_good_0 | classical belt | - | 50 +1 −0.2 | 0.18 ± 0.01 | 37.56 ± 0.03 | 258 ± 200 | 105.6 ± 0.3 | 3 ± 80 | 2457327.73 |
| 2013 RJ109 | detached | - | 50.1 ± 0.6 | 0.35 ± 0.01 | 18.894 ± 0.002 | 246.3 ± 0.2 | 183.648 ± 0.001 | 314.7 ± 0.4 | 2456576.6 |
| 2013 RR98 | detached | - | 50.214 ± 0.005 | 0.2854 ± 0.0001 | 37.76095 ± 5 × 10⁻⁵ | 233.46 ± 0.02 | 62.266 ± 0.0003 | 19.851 ± 0.009 | 2456548.76 |
| 2013 TX171 | resonant | 6:13 | 50.363 ± 0.007 | 0.27665 ± 8 × 10⁻⁵ | 19.5889 ± 0.0003 | 203.66 ± 0.06 | 167.372 ± 0.0003 | 358.63 ± 0.03 | 2456569.66 |
| 2013 RM109 | resonant | 5:11 | 50.83 ± 0.01 | 0.2229 ± 0.0008 | 14.2853 ± 0.0003 | 128.6 ± 0.2 | 261.019 ± 0.001 | 318.08 ± 0.07 | 2456545.69 |

**Table 1.** continued from previous page

| TNO ID | Class | Res. | $a_b$ (AU) | $e_b$ | $i_b$ (deg) | $\omega_b$ (deg) | $\Omega_b$ (deg) | $M_b$ (deg) | Epoch (JD) |
|---|---|---|---|---|---|---|---|---|---|
| 2013 RL109 | outer centaur | - | 51.3 ± 0.5 | 0.42 ± 0.01 | 12.6721 ± 0.0006 | 75 ± 1 | 199.184 ± 0.002 | 25.8 ± 0.1 | 2456575.59 |
| 2015 AS293 | resonant | 4:9 | 51.58 ± 0.02 | 0.3614 ± 0.0004 | 34.4326 ± 0.0001 | 192.68 ± 0.01 | 88.3938 ± 0.0009 | 44.62 ± 0.01 | 2457034.55 |
| s301_good_1002 | detached | - | 51.6 ± 0.2 | 0.344 ± 0.004 | 20.8762 ± 0.0009 | 67.3 ± 0.2 | 191.801 ± -0.001 | 64.2 ± 0.4 | 2456958.66 |
| 2013 TJ172 | detached | - | 51.93 ± 0.005 | 0.28859 ± 9.00E-05 | 27.3788 ± 0.0002 | 248.23 ± 0.02 | 176.5228 ± 0.0002 | 337.21 ± 0.01 | 2456568.79 |
| s200_good_190 | classical belt | - | 52.01 ± 0.02 | 0.2219 ± 0.0004 | 12.1677 ± 0.0004 | 114.02 ± 0.07 | 186.1537 ± 0.0004 | 34.75 ± 0.04 | 2456546.77 |
| 2013 SM102 | detached | - | 52.284 ± 0.006 | 0.25582 ± 8 × 10$^{-5}$ | 11.4072 ± 0.0002 | 226.98 ± 0.03 | 154.0146 ± 0.0006 | 353.27 ± 0.02 | 2456563.62 |
| (529823) 2010 PP81 | detached | - | 52.45 ± 0.03 | 0.2804 ± 0.0002 | 30.7725 ± 0.0001 | 174.9 ± 0.7 | 172.2209 ± -0.0007 | 355.4 ± 0.4 | 2456543.66 |
| s240_good_0 | detached | - | 52.61 ± 0.03 | 0.2738 ± 0.0007 | 28.19157 ± 6 × 10$^{-5}$ | 321.03 ± 0.08 | 38.244 ± 0.001 | 324.41 ± 0.03 | 2457277.51 |
| 2013 SR102 | resonant | 3:7 | 52.85 ± 0.03 | 0.3835 ± 0.0005 | 29.92055 ± 9.00E-05 | 9.76 ± 0.05 | 41.445 ± 0.001 | 298.51 ± 0.04 | 2456565.5 |
| 2013 RO98 | detached | - | 53 ± 5 | 0.3 ± 0.1 | 18.9 ± 0.1 | 90 ± 20 | 292.9 ± 0.2 | 333 ± 8 | 2456540.57 |
| s302_good_31 | resonant | 3:7 | 53.045 ± 0.005 | 0.37415 ± 5 × 10$^{-5}$ | 9.9852 ± 6 × 10$^{-5}$ | 276.32 ± 0.03 | 109.98 ± 0.001 | 0.63 ± 0.01 | 2456544.87 |
| (495297) 2013 TJ159 | resonant | 3:7 | 53.089 ± 0.005 | 0.3173 ± 9.00E-05 | 4.8066 ± 0.0002 | 174.21 ± 0.04 | 165.1691 ± 0.0009 | 11.27 ± 0.02 | 2456546.81 |
| 2014 QG442 | detached | - | 53.7 ± 0.2 | 0.365 ± 0.004 | 30.455 ± 0.001 | 242 ± 0.4 | 95.897 ± 0.002 | 27.63 ± 0.09 | 2456888.92 |
| s200_good_168 | scattering | - | 54.18 ± 0.03 | 0.3984 ± 0.0003 | 18.0483 ± 0.0003 | 81.25 ± 0.02 | 188.996 ± 0.0006 | 29.35 ± 0.006 | 2456548.66 |
| 2014 NB66 | classical belt | - | 55 ± 6 | 0.2 ± 0.1 | 4.69 ± 0.08 | 114 ± 10 | 297 ± 1 | 319 ± 9 | 2457614.78 |
| s200_good_806 | outer centaur | - | 55.06 ± 0.08 | 0.6582 ± 0.0007 | 7.3774 ± 0.0004 | 297.07 ± 0.03 | 290.812 ± 0.004 | 39.04 ± 0.04 | 2456576.59 |
| (495190) 2012 VS113 | detached | - | 55.068 ± 0.002 | 0.30928 ± 2 × 10$^{-5}$ | 26.78573 ± 7 × 10$^{-5}$ | 220.138 ± 0.008 | 171.6043 ± 0.0002 | 1.716 ± 0.004 | 2456243.66 |
| s302_good_124 | resonant candidate | 2:5 | 55.3 ± 0.2 | 0.446 ± 0.003 | 15.068 ± 0.001 | 187.9 ± 0.5 | 170.096 ± 0.004 | 15.6 ± 0.1 | 2456619.71 |
| 2013 RZ108 | resonant | 2:5 | 55.38 ± 0.02 | 0.4525 ± 0.0003 | 13.0319 ± 0.0005 | 333.6 ± 0.2 | 65.597 ± 0.002 | 355.34 ± 0.06 | 2456545.85 |
| 2014 YL50 | resonant | 2:5 | 55.451 ± 0.008 | 0.3251 ± 0.0001 | 29.1463 ± 0.0001 | 234.34 ± 0.04 | 127.3833 ± 0.0005 | 12.7 ± 0.02 | 2457007.68 |
| s12_good_4 | resonant | 2:5 | 55.49 ± 0.04 | 0.359 ± 0.001 | 32.116 ± 0.001 | 338.7 ± 0.4 | 89.063 ± 0.003 | 341.9 ± 0.1 | 2456961.78 |
| 2015 RW245 | outer centaur | - | 56.5 ± 0.1 | 0.531 ± 0.001 | 13.305 ± 0.001 | 19.5 ± 0.6 | 0.39362 ± 7 × 10$^{-5}$ | 356.2 ± 0.1 | 2456578.59 |
| 2014 QT495 | scattering | - | 57.134 ± 0.005 | 0.4739 ± 4 × 10$^{-5}$ | 44.6744 ± 5 × 10$^{-5}$ | 258.73 ± 0.01 | 103.7911 ± 0.0003 | 1.537 ± 0.005 | 2456891.85 |
| s118_good_6 | detached | - | 57.7 ± 0.1 | 0.349 ± 0.004 | 29.964 ± 0.001 | 344 ± 0.7 | 76.068 ± 0.006 | 335.7 ± 0.2 | 2457251.91 |
| 2014 PM82 | detached | - | 58 ± 2 | 0.41 ± 0.04 | 23.8 ± 0.09 | 305 ± 5 | 358.412 ± -0.003 | 23 ± 1 | 2456546.77 |
| 2014 UN225 | detached | - | 59.2 ± 0.04 | 0.3465 ± 0.0007 | 53.1497 ± 0.0001 | 323.42 ± 0.02 | 68.6908 ± 0.0006 | 322.66 ± 0.01 | 2456952.51 |
| s200_good_466 | detached | - | 59.7 ± 0.8 | 0.4 ± 0.01 | 8.393 ± 0.001 | 37 ± 2 | 255.57 ± 0.02 | 21.3 ± 0.4 | 2456548.68 |
| 2014 RS63 | detached | - | 60 ± 2 | 0.39 ± 0.03 | 28.978 ± 0.0009 | 314 ± 7 | 38.91 ± 0.08 | 347 ± 3 | 2456904.69 |
| s200_good_175 | detached | - | 60.3 ± 0.6 | 0.32 ± 0.02 | 11.9467 ± 0.0009 | 213 ± 4 | 186.3212 ± 0.0008 | 336 ± 1 | 2456565.62 |

**Table 1.** continued from previous page

| TNO ID | Class | Res. | $a_b$ (AU) | $e_b$ | $i_b$ (deg) | $\omega_b$ (deg) | $\Omega_b$ (deg) | $M_b$ (deg) | Epoch (JD) |
|---|---|---|---|---|---|---|---|---|---|
| s302_good_160 | detached | - | 60.99 ± 0.03 | 0.3969 ± 0.0003 | 14.1896 ± 0.0002 | 334.77 ± 0.03 | 76.795 ± 0.001 | 345.962 ± 0.009 | 2456568.79 |
| s200_good_272 | detached | - | 61.19 ± 0.06 | 0.416 ± 0.001 | 25.2551 ± 0.2 | 112.3 | 176.1569 ± 0.0001 | 14.94 ± 0.06 | 2456546.7 |
| s301_good_720 | detached | - | 61.53 ± 0.007 | 0.37693 ± 8 × 10$^{-5}$ | 5.4668 ± 0.0002 | 181.18 ± 0.03 | 156.353 ± 0.001 | 11.9 ± 0.01 | 2456568.59 |
| 2013 VQ25 | detached | - | 61.547 ± 0.006 | 0.42287 ± 6 × 10$^{-5}$ | 28.5944 ± 0.0002 | 215.05 ± 0.03 | 153.6874 ± 0.0003 | 3.91 ± 0.01 | 2456888.85 |
| 2014 OQ394 | detached | - | 61.962 ± 0.005 | 0.43965 ± 4 × 10$^{-5}$ | 29.4887 ± 0.0002 | 157.85 ± 0.01 | 186.62073 ± 3 × 10$^{-5}$ | 7.444 ± 0.004 | 2456544.7 |
| (134210) 2005 PQ21 | detached | - | 61.99 ± 0.01 | 0.3931 ± 0.0001 | 6.482 ± 0.0002 | 22.18 ± 0.03 | 316.3337 ± 0.0009 | 4.44 ± 0.01 | 2456540.62 |
| 2014 SP363 | detached | - | 62.409 ± 0.007 | 0.3152 ± 0.0002 | 31.2281 ± 0.0003 | 267.8 ± 0.06 | 145.4485 ± 0.0005 | 341.37 ± 0.02 | 2456982.7 |
| 2010 TJ | detached | - | 62.856 ± 0.006 | 0.365 ± 0.001 | 38.9009 ± 0.03 | 273.92 | 91.2989 ± 0.0004 | 9.9 ± 0.01 | 2456887.91 |
| s13_good_7 | detached | - | 65.14 ± 0.02 | 0.4398 ± 0.0003 | 28.5578 ± 0.04 | 230 | 129.5521 ± 0.0007 | 15.69 ± 0.01 | 2456927.82 |
| 2013 SG102 | outer centaur | - | 65.7 ± 0.4 | 0.62 ± 0.003 | 3.1699 ± 0.0006 | 51.74 ± 0.03 | 199.722 ± 0.002 | 26.77 ± 0.09 | 2456565.62 |
| (480017) 2014 QB442 | detached | - | 66.34 ± 0.01 | 0.4478 ± 0.0001 | 7.2913 ± 0.0002 | 269.32 ± 0.02 | 75.295 ± 0.002 | 12.82 ± 0.006 | 2456568.64 |
| s14_good_4 | detached | - | 67.22 ± 0.01 | 0.3587 ± 0.0003 | 32.478 ± 0.0003 | 287.4 ± 0.1 | 149.5281 ± 0.0009 | 346.89 ± 0.04 | 2456904.9 |
| 2013 SN102 | detached | - | 67.72 ± 0.01 | 0.43879 ± 8 × 10$^{-5}$ | 4.45459 ± 5 × 10$^{-5}$ | 247.36 ± 0.02 | 114.147 ± 0.002 | 1.153 ± 0.007 | 2456564.73 |
| (136199) Eris | detached | - | 67.83 ± 0.03 | 0.4384 ± 0.0004 | 43.993 ± 0.001 | 151.2 ± 0.2 | 35.976 ± 0.001 | 202.7 ± 0.2 | 2456547.89 |
| s301_good_988 | scattering | - | 69 ± 5 | 0.54 ± 0.04 | 11.3 ± 0.1 | 42 ± 2 | 8.69 ± 0.03 | 349.3 ± 0.3 | 2456887.82 |
| 2014 QC442 | detached | - | 69.09 ± 0.02 | 0.5008 ± 0.0001 | 18.99 ± 0.0002 | 45.035 ± 0.007 | 46.6736 ± 0.0003 | 335.608 ± 0.002 | 2456568.64 |
| s12_good_5 | resonant | 2:7 | 69.18 ± 0.08 | 0.5099 ± 0.0009 | 28.2758 ± 0.0003 | 294.5 ± 0.2 | 130.862 ± 0.001 | 349.85 ± 0.05 | 2457003.7 |
| 2015 TW361 | resonant | 2:7 | 69.27 ± 0.01 | 0.46808 ± 6 × 10$^{-5}$ | 16.6857 ± 0.0002 | 331.652 ± 0.009 | 42.2079 ± 0.0001 | 359.956 ± 0.003 | 2456569.66 |
| 2016 SE56 | resonant | 2:7 | 69.5 ± 0.01 | 0.55324 ± 8 × 10$^{-5}$ | 26.7798 ± 0.0001 | 218.799 ± 0.009 | 175.1145 ± 0.0001 | 356.576 ± 0.002 | 2456568.63 |
| 2013 TM172 | detached | - | 69.697 ± 0.009 | 0.4773 ± 6 × 10$^{-5}$ | 12.6083 ± 0.0002 | 352.313 ± 0.009 | 14.8665 ± 0.0007 | 358.455 ± 0.003 | 2456578.63 |
| s302_good_132 | outer centaur | - | 72.24 ± 0.07 | 0.6788 ± 0.0004 | 17.8937 ± 0.0003 | 345.64 ± 0.08 | 63.0945 ± 0.0007 | 356.97 ± 0.01 | 2456568.75 |
| 2016 SS55 | detached | - | 73.15 ± 0.02 | 0.4761 ± 0.0001 | 28.4964 ± 0.0002 | 158.01 ± 0.02 | 182.7956 ± 0.0002 | 16.69 ± 0.003 | 2456568.79 |
| (145480) 2005 TB190 | detached | - | 75.66 ± 0.01 | 0.38939 ± 7 × 10$^{-5}$ | 26.4795 ± 0.0002 | 171.44 ± 0.03 | 180.4517 ± 6 × 10$^{-5}$ | 358.24 ± 0.01 | 2456540.62 |
| 2014 SO350 | resonant | 1:4 | 75.8 ± 0.008 | 0.54352 ± 5 × 10$^{-5}$ | 24.04237 ± 6 × 10$^{-5}$ | 244.161 ± 0.009 | 140.9972 ± 0.002 | 0.866 ± 0.002 | 2456930.76 |
| 2008 UA332 | resonant | 1:4 | 75.83 ± 0.02 | 0.5134 ± 0.0002 | 30.7411 ± 0.0002 | 226.49 ± 0.01 | 109.0105 ± 0.0006 | 18.71 ± 0.002 | 2456915.83 |
| 2014 QV495 | detached | - | 79.56 ± 0.05 | 0.5448 ± 0.0004 | 23.3893 ± 0.0003 | 276.98 ± 0.08 | 69.197 ± 0.001 | 5.75 ± 0.02 | 2456888.83 |
| s11_good_14 | outer centaur | - | 80.8 ± 0.2 | 0.707 ± 0.0008 | 37.132 ± 0.0008 | 215 ± 2 | 167.677 ± 0.0004 | 0.23 ± 0.03 | 2457318.74 |
| 2013 SS102 | scattering | - | 82.4 ± 0.2 | 0.581 ± 0.001 | 19.7477 ± 0.0004 | 10.66 ± 0.05 | 27.9921 ± 0.0005 | 351.778 ± 0.009 | 2456578.73 |
| 2013 RJ109 | detached | - | 83.06 ± 0.08 | 0.526 ± 0.0004 | 14.1241 ± 0.0003 | 145.1 ± 0.04 | 189.4148 ± 0.0004 | 3.52 ± 0.01 | 2456576.6 |
| s12_good_0 | detached | - | 85.75 ± 0.02 | 0.60443 ± 8 × 10$^{-5}$ | 22.90037 ± 8 × 10$^{-5}$ | 277.14 ± 0.02 | 108.86 ± 0.0006 | 0.255 ± 0.004 | 2456931.88 |

**Table 1.** continued from previous page

| TNO ID | Class | Res. | $a_b$ (AU) | $e_b$ | $i_b$ (deg) | $\omega_b$ (deg) | $\Omega_b$ (deg) | $M_b$ (deg) | Epoch (JD) |
|---|---|---|---|---|---|---|---|---|---|
| 2013 RK109 | scattering | - | 89.48 ± 0.02 | 0.61695 $\pm\,7\times10^{-5}$ | 12.8486 ± 0.0001 | 162.35 ± 0.01 | 176.759 ± 0.002 | 6.04 ± 0.002 | 2456544.72 |
| s200_good_122 | scattering | - | 90.01 ± 0.04 | 0.6316 ± 0.0002 | 17.1525 ± 0.0001 | 86.9 ± 0.01 | 192.4784 ± 0.0004 | 8.626 ± 0.001 | 2456538.68 |
| 2013 SO102 | outer centaur | - | 90.5 ± 0.1 | 0.6754 ± 0.0003 | 9.4718 ± 0.0002 | 237.3 ± 0.1 | 141.268 ± 0.002 | 0.86 ± 0.02 | 2456951.77 |
| 2013 SQ102 | resonant candidate | 3:16 | 91.65 ± 0.07 | 0.6162 ± 0.0004 | 29.5484 ± 0.02 | 357.64 ± 0.02 | 14.306 ± 0.0006 | 343.301 ± 0.002 | 2456565.5 |
| (145474) 2005 SA278 | scattering | - | 92.24 ± 0.02 | 0.64155 $\pm\,7\times10^{-5}$ | 16.2753 ± 9.00E- | 277.083 ± 0.007 | 170.3535 ± 0.0004 | 350.2161 ± 0.0008 | 2456268.65 |
| 2014 XY40 | outer centaur | - | 92.9 ± 0.2 | 0.693 ± 0.001 | 28.98795 ± 0.0004 | 336.87 ± 0.02 | 132.529 ± 0.002 | 338.8 ± 0.02 | 2456982.7 |
| s200_good_248 | outer centaur | - | 95.6 ± 0.2 | 0.7396 ± 0.0006 | 13.5922 ± 0.0005 | 79.43 ± 0.02 | 185.2107 ± 0.0003 | 13.434 ± 0.005 | 2456544.67 |
| (437360) 2013 TV158 | outer centaur | - | 97 ± 3 | 0.73 ± 0.01 | 41.1 ± 0.02 | 285.8 ± 0.1 | 191.44 ± 0.02 | 348.56 ± 0.07 | 2456930.78 |
| s302_good_44 | outer centaur | - | 100 ± 6 | 0.71 ± 0.02 | 18.0226 ± 0.0007 | 291.9 ± 0.9 | 177.076 ± 0.005 | 349.31 ± 0.05 | 2456594.67 |
| 2014 SR350 | detached | - | 101 ± 1 | 0.636 $\pm\,6\times10^{-5}$ | 28.76629 $\pm\,6\times10^{-5}$ | 220 ± 0.3 | 35.124 ± 0.003 | 11.9939 ± 0.0003 | 2456886.71 |
| s17_good_0 | detached | - | 104.83 ± 0.03 | 0.5213 ± 0.0002 | 43.1491 $\pm\,6\times10^{-5}$ | 297.15 ± 0.03 | 130.3806 ± 0.0005 | 351.796 ± 0.006 | 2456925.82 |
| 2014 UZ224 | detached | - | 108.8 ± 8 | 0.648 ± 0.003 | 26.7846 ± 0.0003 | 29.4 ± 0.002 | 131 ± 0.2 | 319.4 ± 0.2 | 2456888.92 |
| (437360) 2013 TV158 | detached | - | 111.229 ± 0.006 | 0.67212 $\pm\,2\times10^{-5}$ | 31.14327 $\pm\,8\times10^{-5}$ | 232.106 ± 0.004 | 181.0751 ± 0.0001 | 357.306 ± 0.0005 | 2456575.64 |
| s200_good_624 | outer centaur | - | 111.7 ± 0.4 | 0.7652 ± 0.0009 | 10.2728 ± 0.0002 | 142.63 ± 0.03 | 285.697 ± 0.002 | 348.754 ± 0.006 | 2456564.67 |
| s302_good_209 | scattering | - | 116 ± 5 | 0.68 ± 0.02 | 18.59 ± 0.03 | 272 ± 20 | 120 ± 0.2 | 1 ± 3 | 2456619.75 |
| s200_good_461 | detached | - | 120.84 ± 0.04 | 0.669 ± 0.0001 | 31.6747 ± 0.0002 | 160.02 ± 0.01 | 175.66083 $\pm\,10^{-5}$ | 2.144 ± 0.001 | 2456543.67 |
| 2014 QW495 | scattering | - | 133.5 ± 0.2 | 0.7474 ± 0.0004 | 28.5047 ± 0.0002 | 208.87 ± 0.02 | 75.9361 ± 0.0005 | 2.186 ± 0.002 | 2456898.55 |
| s200_good_520 | scattering | - | 158.4 ± 0.1 | 0.7686 ± 0.0002 | 17.3988 ± 0.02 | 27.39 ± 0.02 | 293.4378 ± 0.0003 | 1.223 ± 0.002 | 2456540.57 |
| s200_good_30 | detached | - | 160 ± 20 | 0.71 ± 0.04 | 4.81 ± 0.05 | 130 ± 40 | 219.4 ± 0.7 | 0 ± 5 | 2457657.63 |
| (508338) 2015 SO20 | scattering | - | 164.8 ± 0.2 | 0.7988 ± 0.0003 | 23.4104 ± 0.0005 | 354.8 ± 0.1 | 33.6343 ± 0.0004 | 359.322 ± 0.009 | 2456545.85 |
| 2016 QV89 | detached | - | 171.64 ± 0.05 | 0.76722 $\pm\,8\times10^{-5}$ | 21.38778 $\pm\,8\times10^{-5}$ | 281.088 ± 0.004 | 173.2158 ± 0.0002 | 354.00533 $\pm\,8\times10^{-5}$ | 2456247.59 |
| (469750) 2005 PU21 | outer centaur | - | 174.6 ± 0.1 | 0.8325 ± 0.0001 | 6.1748 ± 0.0001 | 227.856 ± 0.004 | 192.4938 ± 0.0003 | 355.7945 ± 0.0001 | 2456537.74 |
| s11_good_19 | outer centaur | - | 205.1 ± 0.1 | 0.87333 $\pm\,8\times10^{-5}$ | 26.12526 $\pm\,5\times10^{-5}$ | 298.535 ± 0.003 | 148.5031 ± 0.0003 | 357.44987 $\pm\,5\times10^{-5}$ | 2456888.86 |
| 2016 SG58 | scattering | - | 232.97 ± 0.09 | 0.84931 $\pm\,6\times10^{-5}$ | 13.22082 $\pm\,10^{-5}$ | 296.292 ± 0.0006 | 118.98 ± 0.0003 | 358.8465 ± 0.0003 | 2456568.8 |
| 2013 SL102 | scattering | - | 314.4 ± 0.2 | 0.87871 $\pm\,6\times10^{-5}$ | 6.50488 $\pm\,2\times10^{-5}$ | 265.487 ± 0.008 | 94.732 ± 0.001 | 0.2163 ± 0.0002 | 2456544.71 |
| 2015 BP519 | scattering | - | 449 ± 3 | 0.9215 ± 0.0009 | 54.1107 ± - | 348.06 ± 0.01 | 135.2129 ± 0.0004 | 358.3396 ± 0.0004 | 2456988.83 |
| 2013 RA109 | scattering | - | 462.4 ± 0.4 | 0.9005 $\pm\,8\times10^{-5}$ | 12.39965 $\pm\,4\times10^{-5}$ | 262.91 ± 0.01 | 104.8009 ± 0.0009 | 0.2264 ± 0.0003 | 2456547.89 |



\end{document}